\documentclass[1p]{elsarticle}

\journal{J. Logic. Algebr. Program}

\usepackage{ifxetex}
\usepackage{amsmath}

\ifxetex
\usepackage{fontspec}
\usepackage{unicode-math}
\newfontfamily\glyphs{Symbola}
\else
\usepackage{amssymb}
\usepackage{wasysym}
\usepackage{stmaryrd}
\newcommand\coloneq{:=}
\newcommand\Coloneq{::=}
\def\mdlgwhtcircle{\bigcirc}
\let\mdlgwhtdiamond=\lozenge
\let\mdlgwhtsquare=\square
\let\lBrack=\llbracket
\let\rBrack=\rrbracket
\newcommand\lAngle{\langle\kern-2pt\langle}
\newcommand\rAngle{\rangle\kern-2pt\rangle}
\fi

\newcommand\shepherd{\includegraphics{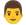}}
\newcommand\wolf{\includegraphics{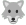}}
\newcommand\goat{\includegraphics{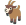}}
\newcommand\cabbage{\includegraphics[scale=.9]{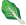}}
\newcommand\apple{\lower1pt\hbox{\includegraphics[scale=.8]{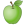}}}
\newcommand\cake{\lower1.5pt\hbox{\includegraphics[scale=.8]{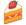}}}

\usepackage{amsthm}
\usepackage{calc}
\usepackage{multirow,booktabs}
\newtheorem{theorem}{Theorem}
\newtheorem{proposition}{Proposition}
\newtheorem{corollary}{Corollary}
\newtheorem{lemma}{Lemma}
\newtheorem{definition}{Definition}

\usepackage{tikz}
\usetikzlibrary{positioning,shapes.geometric,calc,decorations.pathmorphing}
\usepackage{syntax}
\usepackage[colorlinks]{hyperref}
\usepackage[capitalise,noabbrev]{cleveref}

\usepackage{xcolor}
\usepackage{listings}
\usepackage{geometry}

\newcommand*\N{\mathbb{N}}

\newcommand*\kywd[1]{\texttt{\bfseries #1}}
\newcommand*\skywd[1]{{\normalfont\texttt{\color{darkgray}\bfseries #1}}}

\newcommand\idle{{\normalfont\skywd{idle}}}
\newcommand\fail{{\normalfont\skywd{fail}}}
\newcommand\seq{{\normalfont\texttt;}}
\newcommand\disj{{\normalfont\texttt|}}
\newcommand\ifthel[3]{{\normalfont \,#1\, \texttt? \,#2\, \texttt: \,#3}}
\newcommand\cond[2]{{\normalfont\skywd{match}\; #1 \;\skywd{s.t.}\; #2}}

\lstnewenvironment{maudexec}[1][]{\lstset{language={}, #1}}{}

\newcommand*\ao{\,\lower1pt\hbox{@}\,}

\newcommand\subterm{\mathrm{subterm}}
\newcommand\rewcond{\mathrm{rewc}}

\newcommand*\ttrew{\;\text{\tt=>}\;}

\newcommand\ctlAll{\ensuremath{\mathbf{A}\,}}
\newcommand\ctlOne{\ensuremath{\mathbf{E}\,}}
\newcommand\ctlNext{\ensuremath{\mdlgwhtcircle\,}}
\newcommand\ctlAllw{\ensuremath{\mdlgwhtsquare\,}}
\newcommand\ctlEvly{\ensuremath{\mdlgwhtdiamond\,}}	\newcommand\ctlUntil{\ensuremath{\,\mathbf{U}\,}}

\newcommand*\mucAll[1]{\ensuremath{[#1]\,}}
\newcommand*\mucOne[1]{\ensuremath{\langle #1 \rangle\,}}

\newcommand*\sfx[2]{#1 \upharpoonright #2}

\newcommand\opsem{\twoheadrightarrow}
\newcommand\xs{\ensuremath{\mathcal{X\!S}}}
\newcommand\cterm{\mathrm{cterm}}

\usepackage{bussproofs}
\usepackage{algorithm2e}
\usepackage{multicol}

\usepackage[author={Rubén}, color=yellow, icolor=yellow, date={D:20000101000000+01'00'}]{pdfcomment}

\newcommand*\mus[2][\eta]{\lBrack #2 \rBrack_{#1}}
\newcommand*\muss[2][\xi]{\lAngle #2 \rAngle_{#1}}

\newcommand*\ltmsl{\ensuremath{\mathcal M}}

\renewcommand*\skywd[1]{{\normalfont\textsf{\color{darkgray}\bfseries #1}}}

\newcommand*\wprefix[2]{\ensuremath{#1^{\leq\, #2}}}

\crefname{algocf}{Algorithm}{Algorithms}

\lstdefinelanguage{maude}{
	morekeywords=[1]{fmod,endfm,mod,endm,smod,endsm,fth,endfth,th,endth,sth,endsth,view,endv,
		is,from,to,including,inc,protecting,pr,extending,
sort,sorts,subsort,subsorts,op,ops,var,vars,eq,ceq,mb,cmb,rl,crl,strat,strats,sd,csd,if
	},
	morekeywords=[2]{
matchrew,match,amatchrew,amatch,xmatchrew,xmatch,s.t.,or-else,using,by,top,try,idle,fail,not,test,all,one
	},
	alsoletter={.-},
	keywordstyle=[1]{\bfseries},
	keywordstyle=[2]{\color{darkgray}\bfseries},
	morecomment=[l]{***},
	morecomment=[l]{---},
	morestring=[b]"
}

\makeatletter
\def\ps@pprintTitle{\let\@oddhead\@empty
     \let\@evenhead\@empty
     \def\@oddfoot
       {\hbox to \textwidth {\ifnopreprintline\relax\else
        \@myfooterfont \ifx\@elsarticlemyfooteralign\@elsarticlemyfooteraligncenter \hfil\@elsarticlemyfooter\hfil \else \ifx\@elsarticlemyfooteralign\@elsarticlemyfooteralignleft \@elsarticlemyfooter\hfill{}\else \ifx\@elsarticlemyfooteralign\@elsarticlemyfooteralignright {}\hfill\@elsarticlemyfooter \else \normalshape\hfill\begin{tikzpicture}
			\node at (0, 2em) {};
			\node[draw=black!70, fill=black!5, inner sep=5pt, text width=.855\linewidth]{
				Accepted authors' manuscript of the article published in \@journal\ 123 \\
				DOI: \href{https://doi.org/10.1016/j.jlamp.2021.100700}{10.1016/j.jlamp.2021.100700} \hfill License: CC-BY-NC-ND
			};
		\end{tikzpicture} \hfill\fi \fi \fi \fi }
       }\let\@evenfoot\@oddfoot}
\makeatother

\begin{document}

\begin{frontmatter}

\address[ucm]{Facultad de Informática, Universidad Complutense de Madrid, Spain}
\address[itc]{Instituto de Tecnología del Conocimiento, Universidad Complutense de Madrid, Spain}

\cortext[cor1]{Corresponding author}

\title{Strategies, model checking and branching-time \\ properties in Maude}
\author[ucm]{Rubén Rubio\corref{cor1}}
\ead{rubenrub@ucm.es}
\author[ucm,itc]{Narciso Martí-Oliet}
\ead{narciso@ucm.es}
\author[ucm]{Isabel Pita}
\ead{ipandreu@ucm.es}
\author[ucm]{Alberto Verdejo}
\ead{jalberto@ucm.es}

\begin{abstract}
	Rewriting logic and its implementation Maude are a natural and expressive framework for the specification of concurrent systems and logics. Its nondeterministic local transformations are described by rewriting rules, which can be controlled at a higher level using a builtin strategy language added to Maude~3. This specification resource would not be of much interest without tools to analyze their models, so in a previous work, we extended the Maude LTL model checker to verify strategy-controlled systems.
In this paper, CTL* and $\mu$-calculus are added to the repertoire of supported logics, after discussing which adaptations are needed for branching-time properties. The new extension relies on some external model checkers that are exposed the Maude models through general and efficient connections, profitable for future extensions and further applications. The performance of these model checkers is compared.
\end{abstract}


\end{frontmatter}

\section{Introduction}

	Rewriting logic~\cite{rewritingLogic,20years} is a natural and expressive framework for the formal specification and analysis of concurrent systems and logics. Their states are described as terms modulo equations and structural axioms, and their transitions are expressed using rewriting rules. Executing a rewrite system consists of the successive application of a rule in a matching position of the term, both chosen nondeterministically and independently at each step, yielding potentially many evolutions. The spatial and temporal locality of rules is the cornerstone of the natural and simple representation of concurrency, but it is sometimes convenient to tame this nondeterminism and capture the global behavior of the system or other kinds of restrictions. For example, the terms and deduction rules of an inference system can be expressed as a rewrite theory and be proven sound, but only a careful application of these rules will lead to the desired deductions. This idea is enunciated in the Kowalski's motto \emph{Algorithm = Logic + Control}~\cite{kowalski} and developed in the Lescanne's \emph{Rule + Control} approach~\cite{lescanneOrme}, which promotes the separation of the concerns of rules and their control. This is the purpose of strategies, which have been used in formal specification languages like ELAN~\cite{elan}, TOM~\cite{tom}, Stratego~\cite{stratego}, and more recently Porgy~\cite{porgyJournal} for graph rewriting. Unlike strategies usually considered for the $\lambda$-calculus~\cite{barendregt} and abstract rewriting~\cite{allthat,terese}, these are called \emph{programmable strategies} because they are represented syntactically as arbitrary complex programs.

	Maude~\cite{maude,allmaude} is a specification language based on rewriting logic and an interpreter for executing and analyzing its specifications. Strategies have been used in Maude since its beginnings~\cite{clavel96,clavel97} using reflection. However, reflective programs are verbose and difficult to understand for those not used to them, so an object-level strategy language was proposed, prototyped and tested, and finally implemented in Maude 3~\cite{towardsStrategy}. Based on that experience and on earlier languages like ELAN and Stratego, its design puts special emphasis on separating rules from strategies, so that different strategies can be compositionally specified to control the same rewriting system easily.
This new resource for writing formal specifications would be less attractive to Maude users if there were fewer means to work with strategy-controlled models than with standard ones. Hence, we extended the builtin Maude LTL model checker~\cite{maudemc} to support them~\cite{fscd}. It has already been given various applications~\cite{bitmlmc,memstratmc,metatrans}.

	In this paper, we address model checking for strategy-controlled systems against branching-time properties, by first discussing the problem in abstract terms and then particularizing them to strategy-controlled Maude specifications, as we did for linear-time properties. In the general setting, a natural notion of satisfaction arises by considering only the subtree of executions allowed by the strategy when checking branching-time properties, in the same way we consider only the subset of allowed executions when checking linear-time properties. As a practical procedure for model checking according to this definition, we suggest transforming the model to incorporate the restrictions imposed by the strategy, which allows checking virtually any logic supported in the uncontrolled system using its standard algorithms. This is similar to our previous approach for linear-time properties, but this transformation must preserve the branching structure of the original model, for which a certain bisimilarity relation will be required.
In order to check linear-time formulae on systems controlled by the Maude strategy language, we provided it with a small-step operational semantics that determines which are exactly the executions described by a strategy expression and is the base to construct the transformed model where those properties can be checked. However, we will see that the previous transformation is not appropriate and consistent for branching-time properties, and some additional adaptations are required.
Following these principles, we now support logics like CTL, CTL*, and $\mu$-calculus by means of external model checkers. All these logics are implemented in the language-independent model checker LTSmin~\cite{LTSmin}, for which we have developed a plugin with on-the-fly access to the models in the C++ implementation of Maude. Other model checkers are also available as backends like NuSMV~\cite{NuSMV}, the \texttt{pyModelChecking} library~\cite{pymodelchecking}, Spot~\cite{spotomega}, and a custom $\mu$-calculus implementation. All these backends are accessed uniformly using an extensible model-checking tool \texttt{umaudemc} implemented using a \texttt{maude} Python library we have developed~\cite{maude-bindings}. The new model checkers can also be applied to standard Maude specifications for which there was no relevant support for branching-time properties thus far. Moreover, the connections developed for this work can be applied for other purposes, like visualization and other types of analysis.

	\paragraph{Comparison with the workshop paper} This article extends the workshop paper~\cite{btimemc}, introducing the extensible architecture of \texttt{umaudemc} and its connection with other model checkers in addition to LTSmin. The presentation has been improved with further details, the performance of the model checkers has been compared, and related work is discussed.

	\paragraph{Structure of the paper} \Cref{sec:preliminars} reviews some precedents required to follow the rest of the paper. \Cref{sec:idea} describes how model checking is understood for strategy-controlled systems in general. \Cref{sec:example} explains how strategy-controlled systems are specified and model checked in Maude, while~\cref{sec:vending} discusses the specific problems that appear when checking branching-time properties and how they are solved. \Cref{sec:maudesmc} introduces the connections to external model checkers, which are evaluated in~\cref{sec:evaluation}. Related work is reviewed in~\cref{sec:relatedwork}. All the material, including the LTL and branching-time model checkers, their documentation and source code, the examples in this paper and many more, is available online~\cite{stratweb}.

\section{Preliminaries} \label{sec:preliminars}

	Let us recall some basic concepts and notation about strategies, rewriting logic and model checking, which will be extensively used along the paper. Informed readers may safely skip some sections. The Maude strategy language is also introduced together with the small-step operational semantics on which our model checker is based.

\subsection{Strategies and transition systems} \label{sec:strategies}

	A \emph{labeled transition system} (LTS) $\mathcal A = (S, A, R)$ is a set of states $S$, a set of labels or actions $A$, and a labeled binary relation $R \subseteq S \times A \times S$ on the states. Sometimes we consider plain transition systems $\mathcal A = (S, R)$ without transition labels, where $R \subseteq S \times S$ is a usual binary relation. They can be seen as a particular case of labeled transition systems with a single label $\tau$ for all transitions, so most claims about these are valid for those.\footnote{Labeled transition systems can also be embedded in plain transition system by pushing the actions on the states.} Arrows are often used to denote the transition relation, and we write $s \to^a s'$ for $(s, a, s') \in R$ and $s \to s'$ if $s \to^a s'$ for some $a \in A$. We call $s \to s'$ an \emph{execution step} in $\mathcal A$, $s'$ a \emph{successor} of $s$, and an \emph{execution} in $\mathcal A$ is a finite or infinite sequence of states $s_0 \to^{a_1} s_1 \to^{a_2} \cdots \to^{a_n} s_n$ linked by the relation. For convenience, we represent executions as finite words $s_0 a_1 s_1 \cdots a_n s_n \in (S \cup A)^*$ or infinite words $s_0 a_1 s_1 a_2 s_2 \cdots \in (S \cup A)^\omega$ alternating states and actions. In the unlabeled case, actions are dropped from the words $s_0 s_1 \cdots s_n$ or $s_0 s_1 \cdots$. Let $\Gamma^*_{\mathcal A} \subseteq S^*$, $\Gamma^\omega_{\mathcal A} \subseteq S^\omega$ and $\Gamma_{\mathcal A} \subseteq S^\infty$ designate the set of all finite and infinite executions of $\mathcal A$, and the union of both. A subscript $s \in S$ will be added to these sets $\Gamma_{\mathcal A, s}$  to indicate that only executions starting from this state are included.

\subsubsection{Strategies}

	In this general context, strategies have been defined in different ways in the literature~\cite{extstrat}, from which we consider two simple characterizations that we will use almost interchangeably:
\begin{enumerate}
	\item \emph{Extensional strategies} are subsets $E \subseteq \Gamma_{\mathcal A}$ of allowed executions of $\mathcal A$.
	\item \emph{Intensional strategies} are partial functions $\lambda : (S \cup A)^+ \to \mathcal P(A \times S)$ that select the possible next steps to continue an execution $w \in (S \cup A)^+$ based on its history, where the states $(a, s') \in \lambda(ws)$ must always satisfy $s \to^a s'$. In the unlabeled case, this can be simplified to $\lambda : S^+ \to \mathcal P(S)$.
\end{enumerate}
The second definition is widely used in games and other verification logics~\cite{mogaveroJournal,atl}, but the first one is simpler and more expressive. In fact, there is an extensional strategy $E(\lambda) \coloneq \{ s_0 a_1 s_1 \cdots \in (S \cup A)^\infty : (a_{k+1}, s_{k+1}) \in \lambda(s_0 \cdots a_k s_k) \}$ for every intensional strategy $\lambda$, but the converse is not true. Even if an intensional strategy $\lambda_E$ can be defined from an extensional one $E$, $\lambda_E(w) \coloneq \{ (a, s) \in A \times S : wasw' \in E, w' \in (S \cup A)^\infty \}$, some information is lost and the inclusion $E \subseteq E(\lambda_E)$ may be strict.
While an extensional strategy can selectively allow finite executions, in intensional strategies all the prefixes of accepted executions are accepted, because there is no way to indicate that an execution is complete. However, this will not be a problem for model checking, because we usually assume that all executions are nonterminating, or otherwise we complete the finite executions by repeating their last states forever and discard the incomplete ones. Another limitation is that extensional strategies are not necessarily closed while intensional strategies are; for example, the first type may allow executions of the form $a^nb^\omega$ for all $n \geq 0$ but not $a^\omega$, while the second type cannot achieve that. This means that fairness restrictions cannot be represented in the strategy, but this is a reasonable assumption for practical executable strategies, and those restrictions can be treated apart as we suggest for future work.

	In summary, we will represent strategies both intensionally and extensionally, using the most convenient representation in each occasion.

\subsubsection{Execution trees} \label{sec:trees}

Since this work is focused on branching-time properties, we should see the executions of a transition system as a tree instead of as a collection of unrelated execution paths. 
The \emph{execution tree} of $\mathcal A$ from a given state $i \in S$ is the tree whose root is $i$ and whose nodes are states with all their successors as children.
From the graph-theoretic point of view, this can be formalized as the graph $(\Gamma^*_{\mathcal A, i}, \{(ws, wsas') : {s \to^a s'}, w \in (S \cup A)^* \})$, which is acyclic and connected. Each vertex $ws$ consists of the current execution state $s$ and its history $w$ down to the root, whose purpose is disambiguating repeated states that may appear at different branches or depths. However, when tree diagrams are drawn, the history is omitted as it can be inferred from the context.
Notice that a strategy $\lambda$ determines a subtree of that execution tree, namely $(\Gamma^*_{\mathcal A, i}, \{ (w, was) : (a, s) \in \lambda(w) \})$. In the unlabeled case, vertices are only words on states, as usual.

\subsection{Rewriting logic} \label{sec:maude}

	Rewriting logic models change by means of rewriting rules operating on the algebraic terms of an equational logic. These terms are built out of an order-sorted signature given by a set of sorts $S$ and an $S^* \times S$-indexed collection $\Sigma$ of function symbols $f : s_1 \cdots s_n \to s$. Sorts are related by a partial order $s_1 < s_2$ that means subsort inclusion. Given an $S$-sorted family of variables $X$, we consider the set of all terms $T_\Sigma(X)$ on these variables, and substitutions $\sigma : X \to T_\Sigma(X)$ as sort-preserving assignments from variables to terms. A substitution can be recursively extended to a function $\overline\sigma : T_\Sigma(X) \to T_\Sigma(X)$ that replaces all occurrences of the variables in a term, and the composition $\sigma_2 \circ \sigma_1$ of two substitutions is defined $(\sigma_2 \circ \sigma_1)(x) \coloneq \overline{\sigma_2}(\sigma_1(x))$. It satisfies $\overline{\sigma_2 \circ \sigma_1} = \overline{\sigma_2} \circ \overline{\sigma_1}$ in the usual functional sense.\footnote{Funtional composition $f \circ g$ is understood in the order $(f \circ g)(x) = f(g(x))$.} The line over the extension is usually omitted. Terms without variables $T_\Sigma \coloneq T_\Sigma(\emptyset)$ are called \emph{ground terms}.

	In a membership equational logic~\cite{spmel} $(\Sigma, E)$ there are two classes of atomic sentences, conditional \emph{equations} and sort \emph{membership axioms}. Their optional conditions are in turn equations and sort membership formulas that yield Horn clauses of the form:

	\[ 	t = t' \quad \text{if } \bigwedge_i u_i = u'_i \wedge \bigwedge_j v_j : s_j \qquad\qquad
		t : s \quad \text{if } \bigwedge_i u_i = u'_i \wedge \bigwedge_j v_j : s_j \]
where $t : s$ states that $t$ has sort $s$.
These statements induce an equality relation $=_E$ that identifies different terms up to provable equality by $E$. The initial term algebra $T_{\Sigma/E}$ is the quotient of the ground terms $T_\Sigma$ modulo this relation. Although its elements $[t]$ are equivalence classes, we will usually write simply $t$ when no confusion is possible.

	A \emph{rewrite theory} $\mathcal{R} = (\Sigma, E, R)$ is a membership equational logic theory $(\Sigma, E)$ with a set $R$ of rewriting rules. Possibly conditional rewriting rules have the form:
\[ l \Rightarrow r \quad \text{if } \bigwedge_i u_i = u'_i \wedge \bigwedge_j v_j : s_j \wedge \bigwedge_k w_k \Rightarrow w'_k \]
The application of a rule to a term $t$ is the replacement of an instance of $l$ in some subterm of $t$ by $r$ instantiated accordingly, if the condition holds. Conditions of the third type are named \emph{rewriting conditions}, and they are satisfied if the instance of each $w_k$ can be rewritten by the rules in zero or more steps to match $w_k'$. Unlike equations, which are required to be confluent and terminating to make the evaluation of equality decidable and efficient, rules can yield nonterminating and diverging computations.

	Given a set of labels $A$ and an assignment $R \to A$ of a label to each rule in the logic, a rewriting system can be seen as a labeled transition system $(T_{\Sigma/E}, A, \to^1_R)$ whose steps $\to^1_R$ are the single application of a rule to any term in the class and whose actions $A$ are the labels of the rules.\footnote{The transitions of a rewriting system can also be labeled by \emph{proof terms} including all the details of the particular rule application, like its context and substitution~\cite{rewritingLogic}. However, rule labels are enough for our purposes.} Strategies can be considered in this LTS.

	Maude~\cite{maude} is a specification and programming language, where equational and rewrite theories are described compositionally using a notation that does not differ much of the previous mathematical language. These specifications can be executed and analyzed with different commands included in the Maude interpreter and other tools. Further details are available in the Maude manual~\cite{maude} and examples are included in the following sections.

\subsection{The Maude strategy language} \label{sec:slang}

	The Maude strategy language~\cite[\S 10]{maude} is used to control the application of rules by expressing rewriting strategies. Strategy expressions, whose syntax is specified by the $\alpha$ symbol in the grammar below, combine explicit application of rules with a small set of programming constructs.
{\renewcommand\skywd[1]{\texttt{#1}}
\begin{align*}
	\alpha & \,\Coloneq\, \beta \mid \texttt{top(}\beta\texttt{)} \mid \idle \mid \fail \mid \cond PC \mid \alpha \seq \alpha \mid (\alpha \disj \alpha) \mid \alpha \,\texttt* \mid \ifthel\alpha\alpha\alpha \\
		& \;\;\mid\;\;\, \texttt{matchrew} \;P\; \texttt{s.t} \;C \; \texttt{by} \; x \;\texttt{using} \; \alpha , \; \ldots , x \; \texttt{using} \; \alpha \mid \mathit{slabel} \mid \mathit{slabel}\texttt(\vec t\texttt) \\
		& \;\;\mid\;\;\, \alpha \,\texttt+ \mid \alpha \,\texttt! \mid \alpha \;\texttt{or-else}\; \alpha \mid \texttt{test(}\alpha\texttt{)} \mid \texttt{try(}\alpha\texttt{)} \mid \texttt{not(}\alpha\texttt{)} \\
	\beta	& \,\Coloneq\, \mathit{rlabel} \mid \mathit{rlabel}\texttt[\rho\texttt] \mid \mathit{rlabel}\texttt\{\vec\alpha\texttt\} \mid \mathit{rlabel}\texttt[\rho\texttt]\texttt\{\vec\alpha\texttt\} \mid \texttt{all}\end{align*}
}
The first two rows are the essential part of the strategy language, including the rule application strategies under the $\beta$ symbol, since the combinators in the third row can be defined in terms of those of the first two. The meaning of a strategy expression is usually described by the set of terms that its nondeterministic application on a given initial term produces. However, we are also interested in the intermediate states of the strategy-controlled rewriting and in the infinite rewriting sequences allowed by the strategy, which are crucial for model checking. Consequently, we have described the meaning of strategy expressions using a nondeterministic small-step operational semantics~\cite{fscd}. Its steps are defined on \emph{execution states} $q \in \xs$ whose most basic form are pairs $t \ao \alpha_1 \cdots \alpha_n$ where $t$ is the term being rewritten and $\alpha_1, \ldots, \alpha_n$ are the pending strategies to be executed, in that order. However, additional structure will be added to these sets as required by some specific combinators. In any case, the \emph{subject term} being rewritten can be identified from an execution state with the projection $\cterm : \xs \to T_\Sigma(X)$, whose definition on the simpler states is $\cterm(t \ao z) = t$. States $t \ao \varepsilon$ with an empty stack are called \emph{solutions}, since no more work is pending and $t$ can be seen as a result of the strategic computation. Substitutions $\theta : X \to T_\Sigma(X)$ may also be pushed to this execution stack, and they determine the value of the variables in the strategy expressions to their left. In the following, $\theta$ will always refer to the leftmost substitution of the current stack $z$, or to the identity function if there is none. The following are the core combinators of the language:
\begin{itemize}
	\item The rule application strategy $\mathit{rlabel}\texttt[x_1 \,\texttt{<-}\, t_1\texttt, \ldots\texttt, x_n \,\texttt{<-}\, t_n]\texttt\{\alpha_1, \ldots, \alpha_m\}$ executes a single rewrite on the subject term using any rule with label $\mathit{rlabel}$ under some optional restrictions, and produce all possible such rewrites as a result. For rules without rewriting conditions, its semantics is straighforward
\[ t \ao \mathit{rlabel}\texttt[x_1 \,\texttt{<-}\, t_1\texttt, \ldots\texttt, x_n \,\texttt{<-}\, t_n] \; z \to_s t' \ao z \]
The mapping from $x_i$ to $t_i$ between brackets is an optional substitution that is applied to both sides of the rule and its condition before matching, and whose values $t_i$ are previously instantiated with the leftmost substitution $\theta$ in the stack $z$. For a rule with $m$ rewriting conditions, exactly $m$ strategies must be provided between curly brackets to control their evaluation. An additional execution state is introduced to hold a nested state for the rewriting fragment and other information like the matching substitution $\sigma$, the remaining rule condition $C'$ and the remaining strategies, the right-hand side of the rule $r$, the context $c$ where it is applied, and the original term $t$.
\begin{align*}
	 t &\ao \mathit{rl}\hbox{\tt[}x_1 \,\hbox{\tt<-}\, t_1, \ldots, x_n \,\hbox{\tt<-}\, t_n\hbox{\tt]}\hbox{\tt\{}\alpha_1, \ldots, \alpha_k\hbox{\tt\}} \, z \\
	&\to_c \rewcond(p_1 : \sigma(l_1) \ao \alpha_1 \theta, \sigma, C', \alpha_2 \cdots \alpha_k, \theta, r, c ; t) \ao z
\end{align*}
Notice that we have written $\to_s$ in the previous rule but $\to_c$ in the current one, because we want to distinguish which steps are \emph{system} steps that apply rewrite rules to the terms and which are \emph{control} steps that only advance the execution of the strategy. In this case, the term is not actually rewritten until the last rewriting fragment has been solved and the remaining condition $C_0$ is empty or purely equational.
\[ \rewcond(p: t' \ao \varepsilon, \sigma, C_0, \varepsilon, r, c ; t) \ao z \to_s c(\sigma'(r)) \ao z \]
Meanwhile, the substitution $\sigma$ is extended with the values yielded by matching the free variables of the target of each rewriting fragment with their solutions. The next rewriting fragment is then executed after evaluating the equational condition between them.
\[ \begin{array}{l}
	\rewcond(p: t' \ao \varepsilon, \sigma, C_0 \wedge l \ttrew p' \wedge C, \alpha \vec \alpha, \theta, r, c ; t) \ao z \\[2pt]
	\kern1em\to_c \rewcond(p': \sigma'(l) \ao \alpha \, \theta, \sigma', C, \vec \alpha, \theta, r, c ; t) \ao z
\end{array} \]
The subsearch is advanced by applying the semantics recursively. However, both control and system steps in the subsearch are seen as control steps of the whole state, since no rewrite is applied to the subject term. In effect, we define $\cterm(\rewcond(p: q, \sigma, C, \vec \alpha, \theta, r, c; t)) = t$.
\begin{prooftree}
	\AxiomC{$q \to_\bullet q'$}
	\UnaryInfC{$\rewcond(p: q, \sigma, C, \vec \alpha, \theta, r, c ; t) \ao z \to_c \rewcond(p: q', \sigma, C, \vec \alpha, \theta, r, c ; t) \ao z$}
\end{prooftree}

Rules are applied anywhere by default, but matching can be limited to the topmost position by surrounding the strategy with \skywd{top}.\item Tests  check whether the subject term matches the pattern $P$ satisfying the equational condition $C$.
\[ t \ao \hbox{\lstinline[mathescape]{match $P$ s.t. $C$}} \; z \to_c t \ao z \qquad \text{if $t$ matches $\theta(P)$ and satisfies $\theta(C)$}\]
The execution only advances if the test succeeds, and the term is not changed. 
The initial keyword can be changed to \skywd{amatch} to match anywhere, or to \skywd{xmatch} to match with extension for structural axioms (see~\cite[\S\ 4.8]{maude}).
	\item Strategies can be combined with a series of operators like concatenation $\alpha \seq \beta$ that applies $\beta$ on every result of $\alpha$,
\[ t \ao (\alpha \seq \beta) \; z \to_c t \ao \alpha \, \beta \; z,\]
the union $\alpha \disj \beta$ that nondeterministically chooses between $\alpha$ and $\beta$,
\[ t \ao (\alpha \disj \beta) \; z \to_c t \ao \alpha \; z \qquad\qquad t \ao (\alpha \disj \beta) \; z \to_c t \ao \beta \; z,\]
and the iteration $\alpha \texttt*$ that repeatedly executes $\alpha$ a nondeterministic number of times,
\[ t \ao (\alpha \texttt*) \; z \to_c t \ao \alpha \, (\alpha \texttt*) \; z \qquad\qquad t \ao (\alpha \texttt*) \; z \to_c t \ao z.\]
Together with the constants \skywd{idle} and \skywd{fail}, whose results are always the initial or no term at all respectively,
\[ t \ao \skywd{idle} \; z \to_c t \ao z \qquad \text{and no rule for \skywd{fail}},\]
this family of combinators resembles those of regular expressions.
	\item The conditional operator $\alpha \,\texttt{?}\, \beta \,\texttt{:}\, \gamma$ that behaves like $\alpha \seq \beta$ if the condition $\alpha$ produces any result,
\[ t \ao (\alpha \,\texttt{?}\, \beta \,\texttt{:}\, \gamma) \; z \to_c t \ao \alpha \, \beta \; z,\]
but evaluates to the results of the negative branch $\gamma$ if $\alpha$ does not produce any,
\[ t \ao (\alpha \,\texttt{?}\, \beta \,\texttt{:}\, \gamma) \; z \to_c t \ao \, \gamma \; z.\]
This latter rule is only applied if the successors of $t \ao \alpha \; \theta$ are finitely many and none is a solution. In general, we say that a strategy \emph{fails} if it does not produce any result.

	\item The combinator \lstinline[mathescape]{matchrew $P$ s.t. $C$ by $x_1$ using $\alpha_1$, $\ldots$, $x_n$ using $\alpha_n$} allows rewriting selected subterms of the subject term. The subterms matching the distinct variables $x_k$ in the pattern $P$ are rewritten according to the corresponding strategies $\alpha_k$ in parallel, using the rules
\begin{align*}
	t &\ao {\small \skywd{matchrew}\, P \,\skywd{s.t.}\, C \,\skywd{by}\, x_1 \,\skywd{using}\, \alpha_1, \ldots, x_n \,\skywd{using}\, \alpha_n \;} z \\
	 &\to_c \subterm(x_1 : \sigma(x_1) \ao \alpha_1 \, \sigma , \ldots , x_n : \sigma(x_n) \ao \alpha_n \, \sigma ; \sigma_{-\{x_1, \ldots, x_n\}}(P)) \ao z
\end{align*}
for any matching substitution $\sigma$ for $t$ in $\theta(P)$ satisfying $\theta(C)$, and
\begin{prooftree}
	\AxiomC{$q_i \to_\bullet q_i'$}
	\UnaryInfC{$\subterm(\ldots, x_i: q_i, \ldots ; t) \ao z \to_\bullet \subterm(\ldots, x_i: q_i', \ldots ; t) \ao z$}
\end{prooftree}
where $\bullet$ can be either $s$ or $c$, so that system (control) steps in a substate are system (control) steps on the whole state. Since the substates of $\subterm$ rewrite subterms of the subject term, it is natural that their system steps are system steps of the whole state, like rewrites in a subterm are rewrites in the whole term. In fact, the current term of a $\subterm$ state is defined recursively as
\[ \cterm(\subterm(x_1: q_1, \ldots, x_n: q_n; t) \ao z) = t[x_1 / \cterm(q_1), \ldots, x_n / \cterm(q_n)] \]
Finally, the results of a \skywd{matchrew} are the reassembled combinations of their solutions,
	\[ \subterm(x_1: t_1 \ao{} \varepsilon, \ldots, x_n: t_n \ao{} \varepsilon ; t) \ao z \to_c t[x_1 / t_1, \ldots, x_n / t_n] \ao z \]
There are \skywd{amatchrew} and \skywd{xmatchrew} variants like for tests.\item Finally, it is possible to give names to strategy expressions and define them in strategy modules. They should be declared with the signature of the arguments they receive, and with the sort $s$ where they are intended to be applied.
\begin{lstlisting}[mathescape]
strat $\mathit{sname}$ : $s_1$ $\ldots$ $s_n$ @ $s$ .
\end{lstlisting}
Strategies are defined using conditional or unconditional strategy definitions that assign strategy expressions to those names.
\begin{lstlisting}[mathescape]
sd $\mathit{sname}$($p_1$, $\ldots$, $p_n$) := $\alpha$ .
csd $\mathit{sname}$($p_1$, $\ldots$, $p_n$) := $\alpha$ if $C$ .
\end{lstlisting}
Conditions $C$ in strategy definitions share their syntax with equational conditions as explained in~\cref{sec:maude}. These named strategies are called by writing their names followed by a comma-separated list of arguments between parentheses, if any,
\[ t \ao \mathit{slabel}\,(t_1, \ldots, t_n) \, z \to_c t \ao \delta \, \sigma \, z,\]
and the righthand side $\delta$ of any definition whose lefthand side matches that call will be executed with the matching substitution $\sigma$ giving value to its variables. Recursive and mutually recursive definitions are allowed, increasing the expressive power of the language. 
\end{itemize}
There are more combinators that can be derived from the previous, for example, the $\alpha \,\kywd{or-else}\, \beta$ combinator, defined as $\alpha \,\texttt{?}\, \skywd{idle} \,\texttt{:}\, \beta$, that executes $\beta$ only if $\alpha$ fails.

	In order to identify the rewriting paths that are allowed by a strategy, we define the relation ${\opsem} \coloneq {\to_s} \circ {\to_c}^*$ that executes a single system step preceded by as many control steps as required. Clearly, $q \opsem q'$ implies $\cterm(q) \to^1_R \cterm(q')$, so the projections of the executions of this relation are actual rewriting paths. Consider the sets of complete finite and infinite executions of a strategy $\alpha$ from an initial term $t$, where $\opsem^a$ is a $\opsem$ step whose final system transition $\to_s$ applies a rule with label $a$,
\begin{align*}
	\mathrm{Ex}^*(\alpha, t) &\coloneq \{ q_0 a_1 q_1 \cdots a_n q_n : q_0 = t \ao \alpha, q_k \opsem^{a_{k+1}} q_{k+1}, q_n \to_c^* t' \ao \varepsilon, t' \in T_\Sigma(X) \} \\
	\mathrm{Ex}^\omega(\alpha, t) &\coloneq \{ q_0(a_k q_k)_{k=1}^\infty : q_0 = t \ao \alpha, q_k \opsem^{a_{k+1}} q_{k+1} \}
\end{align*}
Only those finite executions ending in a state where a solution can be reached by control steps are included. The extensional strategy denoted by $\alpha$ is then by definition
\[ E(\alpha) \coloneq \bigcup_{t \in T_\Sigma(\emptyset)} E(\alpha, t) \quad \text{ where } E(\alpha, t) \coloneq \cterm(\mathrm{Ex}^*(\alpha, t)) \cup \cterm(\mathrm{Ex}^\omega(\alpha, t)) \]
where the projection $\cterm$ is naturally extended to words (leaving actions untouched) and languages. This strategy is intensional by definition.

	In the Maude interpreter, a command \texttt{srewrite $t$ using $\alpha$} is available for rewriting using a strategy. It shows the results of the strategic rewriting, i.e.\ the last terms of its finite executions $\mathrm{Ex}^*(\alpha, t)$.

\subsection{Model checking} \label{sec:modelchecking}

	\emph{Model checking}~\cite{handbookmc} is an automated verification technique based on an exhaustive examination of the executions of a model to prove or refute properties of its dynamic behavior. Models are usually represented as transition systems whose states are annotated with \emph{atomic propositions}, in terms of which the desired properties are expressed. Such construct receives the name of \emph{Kripke structure} $\mathcal K = (S, \to, I, AP, \ell)$ where $(S, \to)$ is a labeled or unlabeled transition system, $I \subseteq S$ is a set of initial states, $AP$ is the set of atomic propositions, and $\ell : S \to \mathcal P(AP)$ is the \emph{labeling function} that declares which atomic properties are satisfied in each state. For simplicity, it is usually assumed that the transition relation $\to$ is \emph{total}, i.e.\ that every state has a successor, and only infinite executions are considered. If it were not, we could apply the typical \emph{stuttering extension} that repeats the last state of finite executions forever.

	Properties are expressed using \emph{temporal logics} with temporal operators to describe how atomic propositions must occur in time, which are usually separated into two classes~\cite{lamport80}:
\begin{itemize}
	\item \emph{Linear-time} properties are universal properties satisfied by every possible execution of the system. In other words, time is seen as a line where the next step is already determined. The main example is Linear Temporal Logic~\cite{pneuliLTL} (LTL) and its multiple extensions.

	\item \emph{Branching-time} properties reason about the whole execution tree, where multiple futures can be available at any moment. Well-known examples are the Computational Tree Logic~\cite{ctl} (CTL), and the more general CTL*~\cite{ctlstar} that includes both LTL and CTL.
\end{itemize}
Another classification distinguishes \emph{state-based} and \emph{action-based} properties~\cite{actionvsState}, depending on whether formulae refer to propositions of the states or the actions of the models. However, both kinds of properties can be considered together, like in $\mu$-calculus~\cite{kozenMucalc} and the Temporal Logic of Rewriting~\cite{tlr}.

	The semantics of temporal logics is usually defined by means of satisfaction relations $\mathcal K, s \vDash \varphi$. In the case of linear-time properties, the satisfaction of a formula $\varphi$ is reduced to its satisfaction $\mathcal K, \pi \vDash \varphi$ for all the executions $\pi \in \Gamma^\omega_{\mathcal K, s}$ of the system. We could say that a linear-time property accepts or rejects words, while a branching-time one does so with trees. The \emph{model-checking problem} consists of deciding whether this satisfaction relation holds for a given model and property.

	We conclude this section by recalling the notion of bisimulation between Kripke structures~\cite[\S\ 26.3.1]{handbookmc}. Many logics, including the ones implemented in this paper, CTL* and $\mu$-calculus, satisfy the same properties in structures related by bisimulation.

\begin{definition}
	Given two (labeled) Kripke structures $\mathcal K_1 = (S_1, A, R_1, I_1, AP, \ell_1)$ and $\mathcal K_2 = (S_2, A, R_2, I_2, AP, \ell_2)$, a bisimulation is a relation $B \subseteq S_1 \times S_2$ such that if $(s_1, s_2) \in B$ then
\begin{itemize}
	\item $\ell_1(s_1) = \ell_2(s_2)$,
	\item for every action $a$ and state $s_1'$ such that $(s_1, a, s_1') \in R_1$, there is some $s_2' \in S_2$ such that $(s_2, a, s_2') \in R_2$ and $(s_1', s_2') \in B$.
	\item the symmetric condition, with $s_2$ in the role of $s_1$ and so on.
\end{itemize}
\end{definition}
If the Kripke structures are not labeled, the same definition is valid by the usual embedding. Two states $s_1 \in S_1$ and $s_2 \in S_2$ are \emph{bisimilar} if there is a bisimulation relation $B$ such that $(s_1, s_2) \in B$. Two Kripke structures $\mathcal K_1$ and $\mathcal K_2$ as above are \emph{bisimilar} if for every initial state $s_1 \in I_1$ there is a bisimilar initial state $s_2 \in I_2$ and vice versa.
  \section{Model checking strategy-controlled systems} \label{sec:idea}

	Given a (labeled) transition system or Kripke structure $\mathcal K = (S, A, R, I, AP, \ell)$ and a strategy $E \subseteq \Gamma_{\mathcal K}$, we say that $(\mathcal K, E)$ is a strategy-controlled system. In a previous work~\cite{fscd}, we have already discussed what should be understood for the satisfaction of a linear-time property by a strategy-controlled system. Looking at strategies as subsets of the executions of the original model, the notion for linear-time properties is natural and inexorable, properties should only be checked on those allowed executions.
\begin{definition}[{\cite[Definition 2]{fscd}}]
	Let $\varphi$ be a linear-time formula, $(\mathcal K, E) \vDash \varphi$ if $\mathcal K, \pi \vDash \varphi$ for all $\pi \in E$.
\end{definition}
A similar definition could be proposed for branching-time properties, since these are checked on trees and strategies can be seen as subtrees of the execution tree of the original Kripke structure, as explained in~\cref{sec:trees}. However, the definitions of branching-time logics do not usually mention trees explicitly, so we resort to an auxiliary Kripke structure to obtain a clear definition.

\begin{definition}[unwinding] \label{def:unwinding}
	Given a Kripke structure $\mathcal K$ and a strategy $\lambda$, the unwinding $\mathcal U(\mathcal K, \lambda)$ of $\mathcal K$ according to $\lambda$ is the Kripke structure $((S \cup A)^+, A, U, I, AP, \ell \circ \mathrm{last})$ where $(w, a, was) \in U$ if $(a, s) \in \lambda(w)$ and $\mathrm{last}(ws) = s$ for all $w \in (S \cup A)^*$.
\end{definition}

The unwinding of a transition system is a well-known concept~\cite{terese}, but in this case only the executions allowed by the strategy are included. As a graph, it is no other than the execution subtree corresponding to the strategy $\lambda$. In case the underlying transition system is unlabeled, the action labels can be removed from the previous definition. We define the satisfaction of a branching-time property by a strategy-controlled system as the satisfaction in the unwinding:

\begin{definition} \label{def:btime}
	Let $\varphi$ be a branching-time formula, $(\mathcal K, E(\lambda)) \vDash \varphi$ if $\mathcal U(\mathcal K, \lambda) \vDash \varphi$.
\end{definition}

This definition coincides with the previous one on linear-time properties, because the executions of the unwinding projected by $\mathrm{last}$ are exactly those of the strategy. However, it does not have direct practical application since the Kripke structure $\mathcal U(\mathcal K, \lambda)$ is not finite. Fortunately, many logics are invariant by bisimulation, and we can try to find a bisimilar Kripke structure where the standard model-checking algorithms can be applied to decide the satisfaction of $\varphi$. The following theorem claims that this is always possible if the language $E(\lambda)$ is $\omega$-regular, which is a quite general requirement in the context of model checking.

\begin{theorem} \label{thm:regular}
	Given an intensional strategy $\lambda$, there is a finite Kripke structure $\mathcal K'$ bisimilar to $\mathcal U(\mathcal K, \lambda)$ if $E(\lambda)$ is $\omega$-regular. The converse does not hold, but in that case $\ell(E(\lambda))$ is $\omega$-regular.
\end{theorem}

This is the program we will adopt regarding the Maude strategy language: finding a finite (so that model checking is decidable) Kripke structure bisimilar (so that the satisfaction of temporal properties is preserved) to the unwinding (to match Definition 4). In any case, the denotation of strategy expressions in~\cref{sec:slang} via the small-step operational semantics gives all the ingredients for~\cref{def:btime}, so we have already unambiguously established whether a branching-time property is satisfied in a Maude specification with strategies. Whenever this denotation is $\omega$-regular, \cref{thm:regular} tells that the plan depicted at the beginning of the paragraph is a reasonable enterprise. How to find a finite Kripke structure and check properties in practice is discussed in the following sections.

\subsection{Generalization of two logics for strategy-controlled systems} \label{sec:genlogics}

	Now, we provide straightforward generalizations to systems controlled by strategies of the textbook semantics of two temporal logics, CTL* and $\mu$-calculus.
	These definitions agree and confirm the soundness of~\cref{def:btime}, since applying them to a system $(\mathcal K, E(\lambda))$ will be proven equivalent to applying the classical definitions to $\mathcal U(\mathcal K, \lambda)$. 

\subsubsection{CTL*}

CTL*~\cite{ctlstar} is a branching-time temporal logic that extends both LTL and CTL, written using the following grammar:
\begin{align*}
	\Phi & \,\Coloneq\, \bot \mid \top \mid p \mid \neg\, \Phi \mid \Phi \wedge \Phi \mid \Phi \vee \Phi \mid \ctlAll \phi \mid \ctlOne \phi \\
	\phi & \,\Coloneq\, \Phi \mid \neg\, \phi \mid \phi \wedge \phi \mid \phi \vee \phi \mid \ctlNext \phi \mid \ctlEvly \phi \mid \ctlAllw \phi \mid \phi \ctlUntil \phi
\end{align*}
Terms built from $\phi$, called \emph{path formulae}, describe properties of fixed execution paths: $\ctlNext \phi$ tells that the property $\phi$ is satisfied in the next state of the path, $\ctlEvly \phi$ and $\ctlAllw \phi$ say that $\phi$ holds in some or all states of the path respectively, and $\phi_1 \ctlUntil \phi_2$ claims that $\phi_2$ is satisfied in some state and $\phi_1$ holds until then. Terms under the $\Phi$ symbol are called \emph{state formulae} and refer to a state of the transition system, to the atomic properties $p \in AP$ it satisfies, and the paths leaving from it, quantified either universally $\ctlAll \phi$ or existentially $\ctlOne \phi$. LTL is the subset with formulae of the form $\ctlAll \phi$ where $\phi$ does not contain path quantifiers, and the initial $\mathbf{A}$ is left implicit. CTL is the subset in which every path operator is preceded by a quantifier. For example, the CTL formula $\ctlAll \ctlAllw (p \to \ctlOne \ctlEvly p)$ tells that it is possible to reach a state where $q$ holds whenever $p$ holds.

	The semantics of CTL* is usually expressed by a satisfaction relation on states $\mathcal K, s \vDash \Phi$ and on paths $\mathcal K, \pi \vDash \phi$. However, when $\mathcal K$ is controlled by a strategy, a state formula like $\ctlOne \phi$ should not quantify over all paths, but only over those allowed by the strategy. Moreover, this subset of paths may depend not only on the last state but on the whole history of the execution. Consequently, the satisfaction relation for strategy-controlled systems replaces the state $s$ by a (partially consumed) extensional strategy $\mathcal K, E \vDash \Phi$, and path formulae also carry a strategy in addition to the chosen path $\mathcal K, E, \pi \vDash \phi$ where $\pi \in E$.\footnote{The definition of the satisfaction relation $\mathcal K, E, \pi \vDash \phi$ maintains the invariant that $\pi'_0 = \pi_0$ for all $\pi' \in E$. The fourth item in the definition includes $\sfx E{\pi_0}$ so that the invariant holds initially regardless of the input $E$.} To maintain this information in the following recursive definition, we introduce the operation $\sfx E{ws} \coloneq \{ s\pi : ws\pi \in E \}$ that gives the execution paths allowed by a strategy $E \subseteq S^\infty$ to continue from $ws$. Given $\pi = (\pi_k)_{k=0}^\infty$, we denote the suffix from $k$ by $\pi^k = (\pi_{k+n})_{n=0}^\infty$, and the prefix of length $n+1$ by $\wprefix\pi{n} = \pi_0 \cdots \pi_n$. For readability, the initial $\mathcal K$ is omitted.

	\newcommand*\ctldef[2]{\begin{tabular}{p{7em}l}\noindent#1 & iff #2\end{tabular}}

	\begin{enumerate}
		\item \ctldef{$E \vDash p$}{$\forall \, \pi \in E \quad p \in \ell(\pi_0)$}
		\item \ctldef{$E \vDash \neg\, \Phi$}{$E \not\vDash \Phi$}
		\item \ctldef{$E \vDash \Phi_1 \wedge \Phi_2$}{$E \vDash \Phi_1$ and $E \vDash \Phi_2$}
\item \ctldef{$E \vDash \mathbf{E} \, \phi$}{$\exists \, \pi \in E \quad (\sfx E{\pi_0}), \pi \vDash \phi$}
		\item \ctldef{$E, \pi \vDash \Phi$}{$E \vDash \Phi$}
		\item \ctldef{$E, \pi \vDash \neg\,\phi$}{$E, \pi \not\vDash \phi$}
		\item \ctldef{$E, \pi \vDash \phi_1 \wedge \phi_2$}{$E, \pi \vDash \phi_1$ and $E, \pi \vDash \phi_2$}
\item \ctldef{$E, \pi \vDash \ctlNext \varphi$}{$(\sfx E{\pi_0 \pi_1}), \pi^1 \vDash \varphi$}
		\item \ctldef{$E, \pi \vDash \phi_1 \,\mathbf{U}\, \phi_2\,$}{$\exists \, n \geq 0 \kern1ex \sfx E{\wprefix\pi{n}}, \pi^n \vDash \phi_2 \, \wedge \,
			\forall \, 0 \leq k < n \;\; \sfx E{\wprefix\pi{k}}, \pi^k \vDash \phi_1$}
	\end{enumerate}
By the usual equivalences, other operators are indirectly defined. This is a direct generalization of the classical semantic definition~\cite{ctlstar}, and similar variations have appeared in the literature when studying CTL* in the context of tree languages~\cite{ctlRegular} and other extensions of this logic.
The only substantial changes are in (4), where only executions in $E$ are considered, and in (8) and (9), where the strategy argument is updated to the allowed executions from the time point where the recursive relation is evaluated. In fact, this definition coincides with the classical relation if we take $E = \Gamma^\omega_{\mathcal K, s}$.

\begin{proposition} \label{prop:ctl:clas}
	Given a CTL* formula $\Phi$, $\mathcal K, s \vDash \Phi$ iff $\Gamma^\omega_{\mathcal K, s} \vDash \Phi$.
\end{proposition}

As promised in the first lines of this section, we finally claim that checking a CTL* property according to this generalized definition in $\mathcal K$ is the same as doing so with the standard definition on the unwinding or a bisimilar structure, because CTL* is invariant by bisimulation.

\begin{proposition} \label{prop:ctl:unw}
	Given $(\mathcal K, E(\lambda))$ and a CTL* formula $\varphi$, $\mathcal U(\mathcal K, \lambda) \vDash \varphi$ iff $\mathcal K, E(\lambda) \vDash \varphi$.
\end{proposition}

\begin{proposition}[{\cite[Theorem 7.20]{principmc}}]
	Two states, $s_1$ of $\mathcal K_1$ and $s_2$ of $\mathcal K_2$, are bisimilar iff $\mathcal K_1, s_1 \vDash \varphi \iff \mathcal K_2, s_2 \vDash \varphi$ for all CTL* (or for all CTL) formulae $\varphi$.
\end{proposition}

\subsubsection{$\mu$-calculus}

	Modal $\mu$-calculus~\cite{kozenMucalc} is an extension of the Hennessy-Milner logic~\cite{hennessyMilner} with least $\mu$ and great $\nu$ fixed-point operators. It can be used to express edge-aware properties on labeled transition systems using two modalities, $\mucAll a \varphi$ that asserts that all states reachable by an $a$ action satisfy $\varphi$, and $\mucOne a \varphi$ which claims the existence of a successor by $a$ that satisfies $\varphi$. Formulae may contain variables $Z$ bound by fixed-point operators.
\begin{align*}
	\varphi & \,\Coloneq\, \bot \mid \top \mid p \mid Z \mid \neg\, \varphi \mid \varphi \wedge \varphi \mid \varphi \vee \varphi \mid \mucAll a \varphi \mid \mucOne a \varphi \mid \mu Z . \varphi \mid \nu Z . \varphi
\end{align*}
The classical value of a $\mu$-calculus formulae is the set of states in which it holds, written $\mus{\phi}$, where $\eta : \mathrm{Var} \to \mathcal P(S)$ is an assignment of values to the free variables that may appear in nested formulae. This logic is more expressive,\footnote{CTL* formulae can be translated into $\mu$-calculus, but not all $\mu$-calculus can be expressed in CTL*. For example, $\ctlAll \ctlEvly p$ is $\mu Z . (p \vee \mucAll{\tau} Z)$ and $\ctlOne \ctlAllw p$ is $\nu Z . (p \wedge \mucOne{\tau} Z)$, being $\tau$ the only label. However, CTL* cannot express that $p$ is satisfied at all even states $\mu Z. p \wedge \mucAll{\tau} \mucAll{\tau} Z$.} but less intuitive and popular than CTL* and its sublogics. However, model checkers for $\mu$-calculus are available like muCRL2~\cite{mCRL2} and LTSmin~\cite{LTSmin}. As well as the previous logics, $\mu$-calculus is invariant by bisimulation.

\begin{proposition}[{\cite[Theorem 6:10]{handbookmc}}]
	If a state $s_1$ of $\mathcal K_1$ is bisimilar to a state $s_2$ of $\mathcal K_2$ then for every closed $\mu$-calculus formula $\varphi$: $s_1 \in \mus[{\mathcal K_1, \eta_1}]{\varphi}$ iff $s_2 \in \mus[{\mathcal K_2, \eta_2}]{\varphi}$.
\end{proposition}

The following generalization mimics the original definition~\cite[\S 6]{handbookmc}, but the denotation of a formula is a set of trees or strategies $\muss{\varphi} \subseteq \mathcal P(\Gamma_{\mathcal K})$ instead of a set of states. The idea is that a system controlled by strategies $(\mathcal K, E)$ satisfies a $\mu$-calculus formula $\varphi$ iff $E \in \muss{\varphi}$.
A valuation is now $\xi : \mathrm{Var} \to \mathcal P(\mathcal P(\Gamma_{\mathcal K}))$ and $\xi[Z/U]$ is the function $\xi$ with its value for the variable $Z$ replaced by $U$.
\begin{enumerate}
		\newcommand*\mucdef[2]{\begin{tabular}{p{5em}l}#1 & #2\end{tabular}}
		\item \mucdef{$\muss{p}$}{$= \{ T \subseteq \Gamma_{\mathcal K} : \forall \pi \in T \quad p \in \ell(\pi_0) \}$}
		\item \mucdef{$\muss{\neg \varphi}$}{$= \mathcal P(\Gamma_{\mathcal K}) \,\backslash\, \muss{\varphi}$}
		\item \mucdef{$\muss{\varphi_1 \wedge \varphi_2}$}{$= \muss{\varphi_1} \cap \muss{\varphi_2}$}
\item \mucdef{$\muss{Z}$}{$= \xi(Z)$}
		\item \mucdef{$\muss{\mucOne a \varphi}$}{$= \{ T \subseteq \Gamma_{\mathcal K} : \exists \, sa\pi \in T \quad \sfx T {s a \pi_0} \in \muss{\varphi} \}$}
		\item \mucdef{$\muss{\mucAll a \varphi}$}{$= \{ T \subseteq \Gamma_{\mathcal K} : \forall \, sa\pi \in T \quad \sfx T {s a \pi_0} \in \muss{\varphi} \}$}
		\item \mucdef{$\muss{\nu Z . \varphi}$}{$= \bigcup \; \{ F \subseteq \mathcal P(\Gamma_{\mathcal K}) : F \subseteq \muss[\xi{[Z/F]}]{\varphi} \}$}
		\item \mucdef{$\muss{\mu Z . \varphi}$}{$= \bigcap \; \{ F \subseteq \mathcal P(\Gamma_{\mathcal K}) : \muss[\xi{[Z/F]}]{\varphi} \subseteq F \}$}
	\end{enumerate}
For instance, the denotation of an atomic proposition $p$ takes all strategies whose paths satisfy $p$ in their initial terms, instead of all states that satisfy $p$ in the classical definition. Similarly, the modality $\mucOne a \varphi$ takes all strategies with a path that satisfy $\varphi$ after an $a$ transition.
As usual, for the fixpoint in (6) to be well-defined, $\varphi$ must be monotone, so every variable must be under an even number of negations. The semantic definition in (6) and (8) are a consequence of the usual equivalences, for example $\mucAll a \varphi \equiv \neg \mucOne a \neg \, \varphi$.

	The following two results are the counterparts of~\cref{prop:ctl:clas,prop:ctl:unw} for CTL*, and say that the definition is actually a generalization of the classical one, and that it is coherent with the procedure proposed for model checking strategy-controlled systems.

\begin{proposition}
	Given $(\mathcal K, E)$ and a closed $\mu$-calculus formula $\varphi$, $s \in \mus[\mathcal K, \eta]{\varphi}$ iff $\Gamma_{\mathcal K, s} \in \muss[\mathcal K, \xi]\varphi$ for any $\eta$ and $\xi$.
\end{proposition}

\begin{proposition}
	Given $(\mathcal K, E(\lambda))$ and a closed $\mu$-calculus formula $\varphi$, $s \in \lBrack \varphi \rBrack_{\mathcal U(\mathcal K, \lambda), \eta}$ for $s\pi \in E$ iff $E \in \lAngle \varphi \rAngle_{\mathcal K, \xi}$ for any $\eta$ and $\xi$.
\end{proposition}

\section{Specification and model checking in Maude by an example} \label{sec:example}

	In this section and through an example, we explain how strategy-controlled systems can be specified and model checked in Maude. The example is the simple and classical river-crossing puzzle, where a shepherd needs to cross a river carrying a wolf, a goat, and a cabbage. The only means is using a boat that only the shepherd can drive and with room for only one more passenger. Shipping the companions of the shepherd one by one would be a solution, but the wolf would eat the goat and the goat would eat the cabbage as soon as the shepherd leaves them alone. First of all, we should specify the signature of the problem as a functional module.
\begin{lstlisting}
fmod RIVER-DATA is
	sorts Being Side Group River .
	subsorts Being Side < Group .

	ops shepherd wolf goat cabbage : -> Being [ctor] .
	ops left right : -> Side [ctor] .
	op __  : Group Group -> Group [ctor assoc comm] .
	op _|_ : Group Group -> River [ctor comm prec 50] .

	vars G1 G2 : Group .

	op initial : -> River .
	eq initial = left shepherd wolf goat cabbage | right .

	op risky : River -> Bool .
	eq risky(shepherd G1 | G2 wolf goat ) = true .
	eq risky(shepherd G1 | G2 goat cabbage) = true .
	eq risky(G1 | G2) = false [owise] .
endfm
\end{lstlisting}
The characters of the puzzle are declared as constants of sort \texttt{Being} with a multiple operator declaration (\kywd{ops}), and two other constants \texttt{left} and \texttt{right} of sort \texttt{Side} identify both sides of the river. A single being or side tag is a group, since their sorts are subsorts of \texttt{Group}, and more interesting groups can be built with the juxtaposition operator \texttt{\_\_}. Two groups configure a river, one for each border, with the operator \texttt{\_|\_}. Associativity and commutativity are indicated by the \texttt{assoc} and \texttt{comm} attributes of their operator declarations. Groups are associative and commutative because they are sets, and the river is commutative because this will simplify the specification of rules. Finally, the \texttt{initial} position of the puzzle is defined by an equation to the term in which all characters are on the left border. The predicate \texttt{risky} identifies the states in which some being is at risk of being eaten, and it is defined with three equations.\footnote{Equations annotated with the \texttt{otherwise} or \texttt{owise} attribute are executed only after all other equations have failed.}

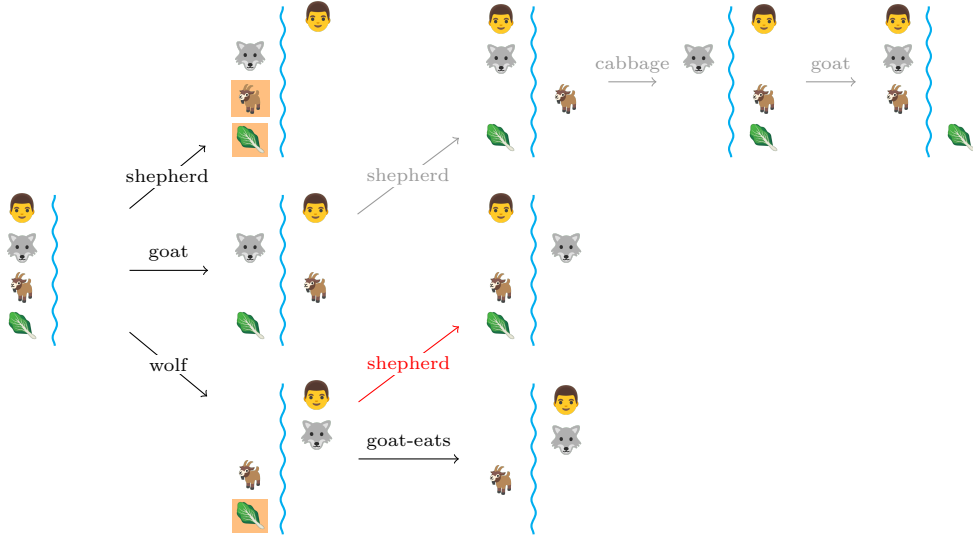
\begin{figure}\centering

\newcommand*\indanger[1]{{\setlength{\fboxsep}{1pt}\colorbox{orange!50}{#1}}}

\newcommand\rcnode[3]{\node (#1) at (#2) {\small\begin{tabular}{cc}
		#3
	\end{tabular}};
	\draw[thick, cyan, decorate, decoration={snake, amplitude=0.8}] ($(#2) + (0, -1)$) -- ($(#2) + (0, 1)$);
}

\leavevmode\kern-1.5em\begin{tikzpicture}
	\rcnode{S0}{-1, 0}{
		\shepherd & \\
		\wolf & \\
		\goat & \\
		\cabbage & \hphantom{\goat} \\
	}

	\rcnode{C1}{2, 2.5}{
		& \shepherd \\
		\wolf & \\
		\indanger{\goat} & \\
		\indanger{\cabbage} & \\
	}

	\rcnode{C2}{2, -2.5}{
		& \shepherd \\
	 	& \wolf \\
		\goat & \\
		\indanger{\cabbage} & \\
	}

	\rcnode{C3}{5.3, -2.5}{
		& \shepherd \\
	 	& \wolf \\
		\goat & \\
	 	& \\
	}

	\rcnode{C4}{5.3, 0}{
		\shepherd &  \\
	 	& \wolf \\
		\goat & \\
	 	\cabbage & \\
	}

	\rcnode{S1}{2, 0}{
		& \shepherd \\
		\wolf & \\
		& \goat \\
		\cabbage & \\
	}

	\rcnode{S2}{5.3, 2.5}{
		\shepherd &  \\
		\wolf & \\
		& \goat \\
		\cabbage & \\
	}

	\rcnode{S3}{7.9, 2.5}{
		 & \shepherd  \\
		\wolf & \\
		& \goat \\
		& \cabbage \\
	}

	\rcnode{S4}{10.5, 2.5}{
		\shepherd &   \\
		\wolf & \\
		\goat &  \\
		& \cabbage \\
	}

	\draw[->] (S0) edge node[fill=white, inner sep=1pt] {\scriptsize shepherd} (C1);
	\draw[->] (S0) edge node[above] {\scriptsize goat} (S1);
	\draw[->] (S0) edge node[fill=white, inner sep=1pt] {\scriptsize wolf} (C2);

	\draw[->, black!40] (S1) edge node[fill=white, inner sep=1pt] {\scriptsize shepherd} (S2);
	\draw[->, black!40] (S2) edge node[above] {\scriptsize cabbage} (S3);
	\draw[->, black!40] (S3) edge node[above] {\scriptsize goat} (S4);

	\draw[->] (C2) edge node[above] {\scriptsize goat-eats} (C3);
	\draw[->, red] (C2) edge node[fill=white, inner sep=1pt] {\scriptsize shepherd} (C4);

\end{tikzpicture}
\caption{Partial rewriting tree for the river-crossing puzzle.} \label{fig:rivertree}
\end{figure}

	On top of this functional module, the possible moves of the game are specified using rules. The system module \texttt{RIVER} imports the functional module \texttt{RIVER-DATA} and defines a rule to cross the river with each character, and two more rules \texttt{wolf-eats} and \texttt{goat-eats} that make the mentioned animal eat its colleague one trophic level below.
\begin{lstlisting}
mod RIVER is
	protecting RIVER-DATA .

	vars L R : Group .

	rl [alone]   : shepherd L | R => L | R shepherd .
	rl [wolf]    : shepherd wolf L | R  => L | R shepherd wolf .
	rl [goat]    : shepherd goat L | R => L | R shepherd goat .
	rl [cabbage] : shepherd cabbage L | R 
	                             => L | R shepherd cabbage

	rl [wolf-eats] : wolf goat L | R shepherd 
	                   => wolf L | R shepherd .
	rl [goat-eats] : goat cabbage L | R shepherd 
	                      => goat L | R shepherd .
endm
\end{lstlisting}
The execution of these rules does not guarantee that the rules of the game are respected, since escaping from a risky state without applying \texttt{wolf-eats} or \texttt{goat-eats} is possible, as shown in~\cref{fig:rivertree}. This suggests that the eating rules must be applied eagerly before any movement rule is executed again, for which strategies will be helpful.\footnote{In a previous Maude specification of the river-crossing puzzle~\cite{playingMaude}, eating actions are written as equations so that they are applied eagerly by the Maude engine before the moving rules. However, this yields a rewrite theory where rules and equations are not coherent. A rewrite theory is \emph{coherent} if for any term $t$ rewritten by a rule to a term $t'$, its canonical form $u$ modulo equations and axioms can be rewritten to a term $u'$ that is equationally equivalent to $t'$, see \cite[\S 5.3]{maude}. Coherence is assumed by Maude, which reduces terms to their canonical forms before applying a rule, not to miss any rewrite.}

	This specification can already be executed within Maude. For instance, the \texttt{search} command finds terms matching a given pattern on the rewriting tree. We can use it to find out whether the goal position of the game can be reached.
\begin{maudexec}
Maude> search initial =>* left | right shepherd wolf goat cabbage  .

Solution 1 (state 32)
states: 33  rewrites: 64
empty substitution

No more solutions.
states: 36
\end{maudexec}
The answer is affirmative, but we cannot be sure whether this state has been reached according to the rules of the game. In fact, the path that the search algorithm has followed to reach the goal position visits risky states, as can be seen using the \texttt{show path} command with the state number that appears next to the solution.
\begin{maudexec}
Maude> show path 32 .
state 0, River: right | left shepherd wolf goat cabbage
===[ rl ... [label wolf] . ]===>
state 2, River: left goat cabbage | right shepherd wolf
===[ rl ... [label alone] . ]===>
state 8, River: right wolf | left shepherd goat cabbage
...
state 32, River: left | right shepherd wolf goat cabbage
\end{maudexec}
The second state in the path is a dangerous position where the \texttt{goat} can eat the \texttt{cabbage}, but this is not actually done in the third one. However, there may be other legitimate paths to the goal.

\subsection{Controlling the system with strategies}

	In order to avoid that situation and enforce the game rules, various strategies will be defined in a strategy module \texttt{RIVER-STRAT} including \texttt{RIVER}.
\begin{lstlisting}
smod RIVER-STRAT is
	protecting RIVER .

	var G : Group .

	strats oneCrossing eating cross&eat @ River .

	sd oneCrossing := alone | wolf | goat | cabbage .
	sd eating      := wolf-eats | goat-eats .
	sd cross&eat   := eating or-else oneCrossing .

	strats eagerEating safe @ River .

	sd eagerEating := (match left | right shepherd wolf
	     cabbage goat) ? idle : (cross&eat ; eagerEating) .
	sd safe := (match left | G) ? idle
	            : (oneCrossing ; not(eating) ; safe) .
endsm
\end{lstlisting}
The auxiliary strategies \texttt{oneCrossing} and \texttt{eating} apply any of the four movement rules and any of the two eating rules, respectively, and \texttt{cross\&eat} applies either one according to the rules of the game, crossing only if eating is not possible. \texttt{eagerEating} is a recursive strategy that repeats this step forever or until the goal is found. \texttt{safe} is more restrictive and avoids visiting risky states by discarding all paths where eating is possible with $\skywd{not}\texttt(\alpha\texttt) \equiv \alpha \,\texttt?\, \skywd{fail} \,\texttt:\, \skywd{idle}$. For example, the bottom branch of~\cref{fig:rivertree} will be allowed by \texttt{eagerEating} but not by \texttt{safe}. Executing \texttt{safe} still requires visiting the risky state at the bottom to find out whether it is actually risky, but this execution path is discarded as if the state were never visited.

	Now, we can ask whether the goal can be properly reached by evaluating \texttt{eagerEating} from the initial state with the \texttt{srewrite} command. The answer is positive.
\begin{maudexec}
Maude> srew initial using eagerEating .

Solution 1
rewrites: 74
result River: left | right shepherd wolf goat cabbage

No more solutions.
rewrites: 74
\end{maudexec}

\subsection{Preparing the specification for model checking}

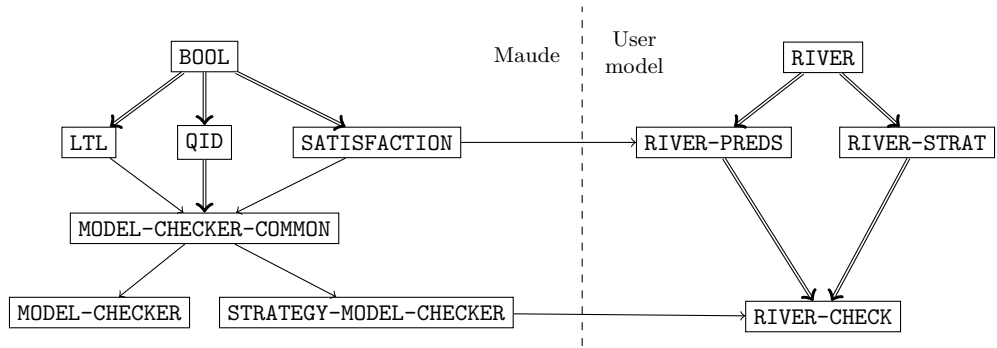
\begin{figure}[t]\centering
	\begin{tikzpicture}[node distance=.7cm and .8cm, scale=0.9, every node/.style={scale=0.9}]
		\node[draw] (SAT) {\ttfamily SATISFACTION};
		\node[draw, left=of SAT] (QID) {\ttfamily QID};
		\node[draw, left=of QID] (LTL) {\ttfamily LTL};

		\node[draw, above=of QID] (BOOL) {\ttfamily BOOL};

		\node[draw, below=of QID] (CC) {\ttfamily MODEL-CHECKER-COMMON};

		\node[draw, below left=.7cm and -4.5em of CC] (C) {\ttfamily MODEL-CHECKER};
		\node[draw, below right=.7cm and -4.5em of CC] (SC) {\ttfamily STRATEGY-MODEL-CHECKER};

		\node[draw, right=2.3cmof SAT] (MP) {\ttfamily RIVER-PREDS};
		\node[right=.3cm of MP] (anchor) {\vphantom{M}};
		\node[draw, right=.1cm of anchor](SM) {\ttfamily RIVER-STRAT};
		\node[draw, above=of anchor](M) {\ttfamily RIVER};
		\node[draw, below=1.88cm of anchor] (MC) {\ttfamily RIVER-CHECK};

		\draw[->] (SAT) -- (MP);
		\draw[double, ->] (M) -- (MP);
		\draw[double, ->] (M) -- (SM);
		\draw[double, ->] (MP) -- (MC);
		\draw[double, ->] (SM) -- (MC);
		\draw[double, ->] (QID) -- (CC);
		\draw[->] (LTL) -- (CC);
		\draw[->] (SAT) -- (CC);
		\draw[->] (CC) -- (C);
		\draw[->] (CC) -- (SC);
		\draw[->] (SC) -- (MC);
		\draw[double, ->] (BOOL) -- (LTL);
		\draw[double, ->] (BOOL) -- (QID);
		\draw[double, ->] (BOOL) -- (SAT);

\draw[dashed] (3cm, -3cm) -- +(0, 5cm);
		\node[anchor=east] at (2.8cm, 1.3cm) {\small Maude};
		\node[anchor=west] at (3cm, 1.3cm) {\small \begin{tabular}cUser\\model\end{tabular}};
	\end{tikzpicture}
	\caption{Structure of the strategy model-checker modules.} \label{fig:smcheck}
\end{figure}

	The last step for checking the model we have just specified, either with the previous model checkers or with those proposed in this paper, is the declaration of its atomic propositions on which temporal properties will be based. The Maude manual~\cite[\S 12]{maude} describes the required steps for strategy-free specifications, and the procedure does not hardly change for strategy-aware ones. It involves a few modules included in the \texttt{model-checker.maude} file shipped with the official and with our extended distribution of Maude, as shown in~\cref{fig:smcheck}.

	Following with the example, the river-crossing puzzle is specified in the \texttt{RIVER} and \texttt{RIVER-STRAT} modules. First, we have to extend the system module \texttt{RIVER} by declaring some atomic propositions as Maude symbols and defining when they are satisfied.\footnote{The declaration of the atomic propositions could have also been done in an extension of \texttt{RIVER-STRAT}. However, as a general principle, it is recommended not to include other content in strategy modules than strategy declarations and definitions, to emphasize their distinct concerns.}
\begin{lstlisting}
mod RIVER-PREDS is
	protecting RIVER .
	including SATISFACTION .

	subsort River < State .
	ops goal risky death : -> Prop [ctor] .

	var  R    : River .
	var  B    : Being .
	vars G G' : Group .

	eq left | right shepherd wolf goat cabbage |= goal = true .
	eq R |= goal = false [owise] .
	eq G cabbage | G' goat |= death = false .
	eq G cabbage goat | G' |= death = false .
	eq R |= death = true [owise] .
	eq R |= risky = risky(R) .
endm
\end{lstlisting}
Atomic propositions must be declared within the sort \texttt{Prop} introduced by the \texttt{SATISFACTION} module of the model checkers' infrastructure. This module also declares a satisfaction symbol \texttt{\_|=\_ : State Prop -> Bool} that should be defined with equations for every state and atomic proposition. The states of the specified system must belong to the sort \texttt{State} appearing in the signature of \texttt{\_|=\_}, for what we have declared \texttt{River} as a subsort of \texttt{State}. Three propositions have been defined: \texttt{goal} that holds on the goal position, \texttt{death} that is only false when all eatable characters are in the scene, and \texttt{risky} that labels the risky states.

	Finally, the \texttt{STRATEGY-MODEL-CHECKER} module, which gives access to the model checker, should be included in a new strategy module incorporating the property specification in \texttt{RIVER-PREDS} and the strategy specification in \texttt{RIVER-STRAT}.
\begin{lstlisting}
smod RIVER-CHECK is
	protecting RIVER-STRAT .
	protecting RIVER-PREDS .
	including STRATEGY-MODEL-CHECKER .
	including MODEL-CHECKER .
endsm
\end{lstlisting}
At this point, given an initial term $t$ and a strategy expression $\alpha$, the Kripke structures that represent the strategy-free model and the strategy-aware model in \texttt{RIVER-CHECK} are completely specified. In the standard case, the model is the rewrite graph reachable from the initial term $t$ where atomic propositions are evaluated using the \texttt{\_|=\_} symbol, i.e.
\[  \mathcal M \coloneq (T_{\Sigma/E}, A, (\to^1_{\mathcal R})^\bullet, \{t\}, AP_{\Pi}, L_{\Pi}) \]
where the actions $A$ are the labels assigned to the rules, the atomic propositions $AP_\Pi$ are the ground instances of \texttt{Prop} symbols, and $L_\Pi$ maps a state to the set of those terms that are reduced to the term \texttt{true} by the equations. The relation $(\to^1_{\mathcal R})^\bullet$ is the one-step rule application $\to^1_{\mathcal R}$ where deadlock states are added a self loop to implement the stutter extension and work with infinite executions only.
In the strategy-aware case, the model used by the LTL model checker guarantees that properties $\varphi$ are only checked on the executions allowed by the strategy $E(\alpha, t)$, i.e.\ $\mathcal M, \pi \vDash \varphi$ for all $\pi \in E(\alpha, t)$. This Kripke structure is given by the graph of the small-step operational semantics described in~\cref{sec:slang},
\[ \mathcal M_\alpha \coloneq (\xs, A, \opsem, \{t \ao \alpha\}, AP_\Pi, L_\Pi \circ \mathrm{cterm}) \]
The relation $\opsem$ is also extended, but only on complete finite executions, i.e.\ only those leading to a solution by control transitions are added the self loop.\footnote{An execution state of the semantics may at the same time lead to a solution and to a new rewrite, so adding a loop is not always safe and these states must be duplicated.}

	These are the models internally used by the Maude LTL model checker and by our previous extension for strategy-controlled systems. For using the latter, the \texttt{STRATEGY-MODEL\-CHECKER} module declares a symbol \texttt{modelCheck($s$, $\varphi$, '$\mathit{name}$)} whose equational reduction invokes the verification of the LTL property $\varphi$ from the initial state $s$ controlled by the strategy whose name is $\mathit{name}$. The property $\varphi$ is expressed as a Maude term whose syntax is specified in the \texttt{LTL} module in \texttt{model-checker.maude}. For instance, we can check the LTL properties $\ctlAllw (\mathit{risky} \rightarrow \ctlNext \mathit{death})$ with \texttt{eagerEating} and $\ctlEvly \mathit{goal}$ with \texttt{safe}. The first one is satisfied, because eating rules are applied eagerly, but the second is not and a counterexample is shown.

\begin{figure}
\includegraphics[width=\linewidth]{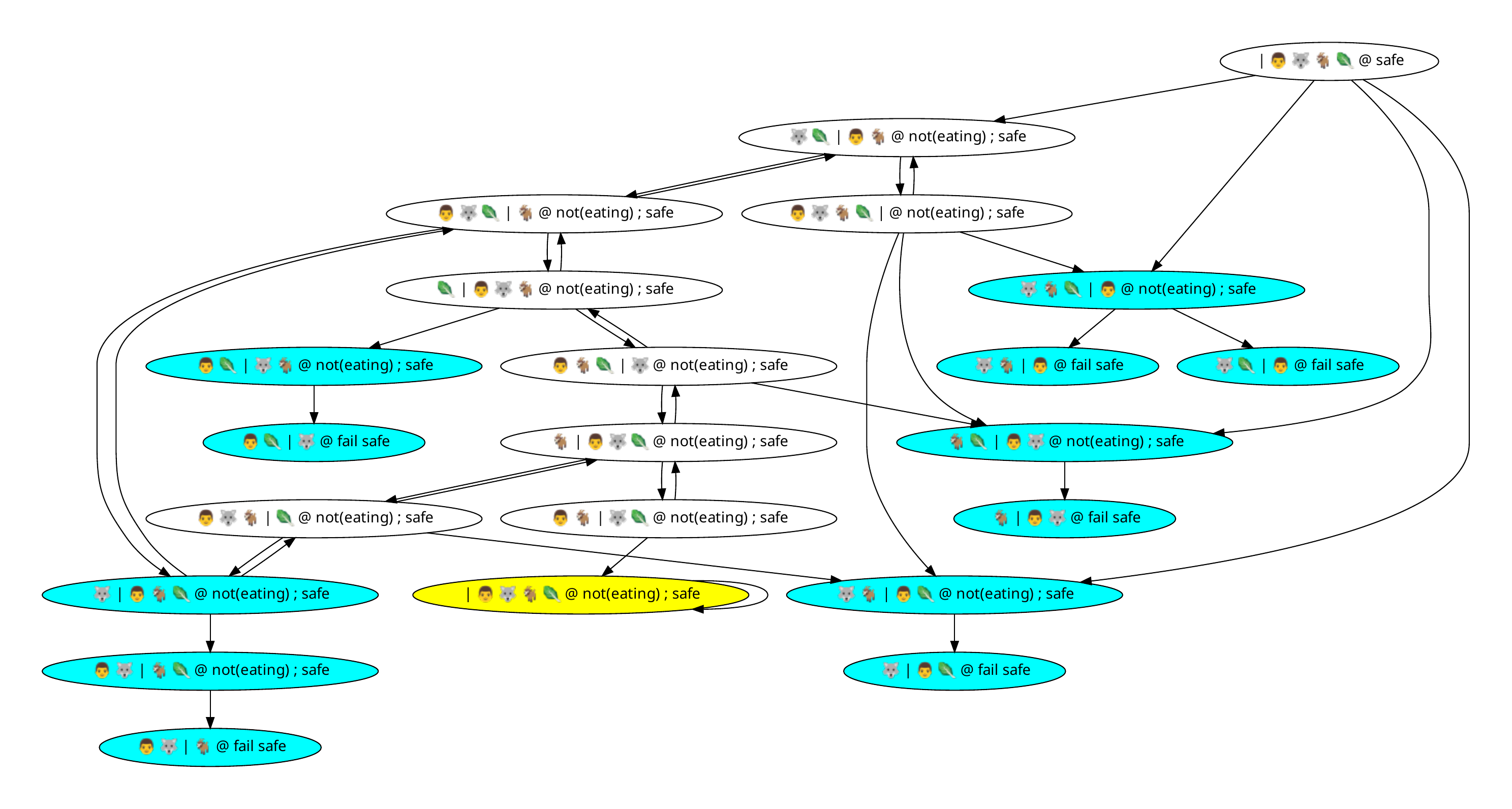}
\caption{Model graph for the \texttt{safe} strategy.} \label{fig:safe}
\end{figure}

\begin{maudexec}
Maude> red modelCheck(initial, [] (bad -> O death), 'eagerEating) .
rewrites: 123
result Bool: true
Maude> red modelCheck(initial, <> goal, 'safe) .
reduce in RIVER-CROSSING-SCHECK : modelCheck(initial, <> goal, 'safe) .
rewrites: 36
result ModelCheckResult: counterexample(
  {right | left shepherd wolf goat cabbage,'goat}
  {left wolf cabbage | right shepherd goat,'alone}
  {right goat | left shepherd wolf cabbage,'wolf}
  {left cabbage | right shepherd wolf goat,'goat}
  {right wolf | left shepherd goat cabbage,'cabbage},
  {left goat | right shepherd wolf cabbage,'alone}
  {left shepherd goat | right wolf cabbage,'alone})
\end{maudexec}
This counterexample, not being as short as possible, shows that it is always possible to repeat movements in a loop. The complete graph for the \texttt{safe} strategy is shown in~\cref{fig:safe}, although the C++ implementation does not explicitly retain the strategy continuation of the semantics. The standard model checker can be used at the same time and it has a similar interface, where the strategy name is obviously omitted. For example, even in the uncontrolled system, the property $\ctlAllw (\mathit{death} \rightarrow \ctlAllw \neg\, \mathit{goal})$ is satisfied.
\begin{maudexec}
Maude> red modelCheck(initial, [] (death -> [] ~ goal)) .
rewrites: 156
result Bool: true
\end{maudexec}

	In principle, using these Kripke structures, we will be able to check properties in no matter which logic. Without strategies, the Kripke structure directly represents the genuine rewrite graph, so there is no problem on applying other model-checking algorithms. However, our transformed strategy-aware structure only guarantees that its nonterminating executions coincide with the denotation of the strategy, but this is not enough. Looking at~\cref{fig:safe}, we see that some states marked in blue do not lead to any solution or infinite execution. They are the states where \lstinline[mathescape]{not(eating)} has failed and where \texttt{eating} has been applied to figure it out on the fly. The depth-first search of the automata-theoretic approach used in the Maude LTL model checker ignores them automatically, since no cycle can be found through them, but algorithms for branching-time properties and tableau-based methods for LTL do not enjoy this property. These \emph{failed states}
 can be safely removed when backtracking on the model generation or using an additional search, but there is another more serious problem that we explain in the following section.

 \section{Strategies and branching-time properties in Maude} \label{sec:vending}

\begin{figure}\centering
	\begin{tikzpicture}[sibling distance=3cm, ->>]
		\node at (0, 0) {\texttt{€€ []} @ $\alpha$}
			child { node[minimum width=57pt] {\texttt{€ [€]} @ apple}
				child { node {\texttt{€ \apple{} []} @ $\varepsilon$ } }}
			child { node[minimum width=72pt] {\texttt{€ [€]} @ put1 ; cake}
				child { node {\texttt{[€€]} @ cake}
					child { node {\texttt{\cake{} []} @ $\varepsilon$} } } };

		\node at (6.5, 0) {\texttt{€€ []} @ $\beta$}
			child { node {\texttt{€ [€]} @ apple | put1 ; cake}
				child { node[minimum width=45pt] {\texttt{€ \apple{} []} @ $\varepsilon$} }
				child { node[minimum width=50pt] {\texttt{[€€]} @ cake}
					child { node {\texttt{\cake{} []} @ $\varepsilon$} } }
			};
	\end{tikzpicture}
	\caption{Strategy rewrite graph for $\alpha$ and $\beta$.} \label{fig:vending}
\end{figure}
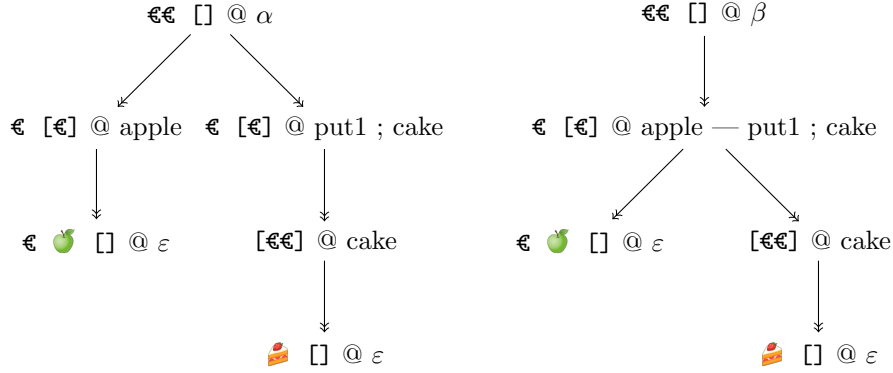

	The transition system yielded by the semantics is not ready for model checking branching-time properties, as seen in the previous section. However, the main reason is that states which are logically the same in the underlying system may be seen as distinct states due to the strategy continuation they hold, changing the tree structure of the model and making it depend on syntactical aspects of the strategies. We will illustrate this problem with an example of a simple vending machine:
\begin{lstlisting}
mod VENDING-MACHINE is
	sorts Soup Thing Machine .
	subsort Thing < Soup .

	ops e a c : -> Thing [ctor] .
	op _[_] : Soup Soup -> Machine [ctor] .

	op empty : -> Soup [ctor] .
	op __ : Soup Soup -> Soup [ctor asoc comm id: empty] .

	vars O I : Soup .

	rl [put1]  : O e [I]     => O   [I e] .
	rl [apple] : O   [I e]   => O a [I] .
	rl [cake]  : O   [I e e] => O c [I] .
endm
\end{lstlisting}
The vending machine is a term \lstinline[mathescape]{$O$ [$I$]} where $O$ represents the belongings of its user and $I$ the content of its internal coin box. The machine can receive one euro coin \texttt{e} with the rule \texttt{put1}, and sells apples \texttt{a} and cakes \texttt{c} for one and two euros respectively.
Let us consider $\alpha \equiv \texttt{put1 ; apple | put1 ; put1 ; cake}$ and $\beta \equiv \texttt{put1 ; (apple | put1 ; cake)}$. These two different strategy expressions are essentially the same, because their abstract denotations coincide $E(\alpha) = E(\beta)$, so the vending machine must satisfy the same properties whether controlled by $\alpha$ or $\beta$ according to~\cref{def:btime}. Intuitively, these strategies can be identified with the plans of a person using the machine, where $\alpha$ has already decided which item to buy before inserting any coin, and $\beta$ delays the choice until the first coin is inserted. An external observer looking at the user interaction with the machine will not be able to distinguish when this choice has been made, it is not part of the \emph{observable behavior}, and so it should be irrelevant for any property considered. This principle would not be obeyed if we applied standard algorithms on $\mathcal M_\alpha$ as~\cref{fig:vending} shows. There, we can see the execution trees by the $\opsem$ relation from an initial configuration with two coins \texttt{e e [empty]} using both $\alpha$ (left) and $\beta$ (right). Disregarding the strategy continuations after the \texttt{@} sign, i.e.\ projecting the nodes by the $\cterm$ function, we obtain rewriting trees where terms are connected by one-step rule rewrites. However, the tree for $\alpha$ cannot be considered a subtree of the execution tree of $(T_{\Sigma/E}, \to^1_R)$ because it contains repeated children. In any case, the branching structures of the execution trees for $\alpha$ and $\beta$ and of their projections are manifestly different, and so they can be distinguished by branching-time temporal properties, as the CTL property $\ctlAll \ctlNext \ctlOne \ctlEvly \mathit{hasCake}$ attests.
\begin{lstlisting}
mod VENDING-MACHINE-PREDS is
	protecting VENDING-MACHINE .
	including SATISFACTION .

	sort Machine < State .
	op hasCake : -> Prop [ctor] .

	vars I O : Soup .
	eq O c [I] |= hasCake = true .
	eq O   [I] |= hasCake = false [owise] .
endm
\end{lstlisting}
In effect, in the immediate successors of the root of the $\alpha$ tree the full path is already chosen, and in the left one no cake is ever bought. On the contrary, there is only one immediate successor of the initial state for $\beta$, where we can still choose the right branch to get the cake.

	The ambiguity on the satisfaction of the atomic property by the strategy $E(\alpha) = E(\beta)$ should be avoided. In this example, the problem would be solved if the two successors of the root in the execution tree for $\alpha$ were combined into a single state, whose projection will be well-defined since they share the same term. Merging successors with a common base term is a general solution to the problem that can be applied locally, solves the ambiguity, and produces a Kripke structure bisimilar to the unwinding of the strategy as desired.
The following definition formalizes this construction and the removal of failed states discussed in the previous section. Remember that a state is valid
\[ \mathrm{valid}(q) \coloneq \exists\, t \in T_\Sigma \quad q \to_{s,c}^* t \ao \varepsilon \quad\vee\quad \exists\, (q_n)_{n=1}^\infty \quad q \opsem q_1 \opsem q_2 \opsem \cdots \]
if a solution or a nonterminating execution can be followed from it.

\begin{definition} \label{def:oprima}
	Given a strategy expression $\alpha$ and $t \in T_\Sigma$, we define the Kripke structure $\ltmsl'_\alpha \coloneq (\xs', A, [\opsem]', \{ \{ t \ao \alpha \} \}, \mathrm{AP}_\Pi, L_\Pi \circ \cterm)$ where
\[ \xs' = \{ Q \subseteq \mathcal P(\xs) : \exists\, t \in T_\Sigma \quad \forall q \in Q \;\; \cterm(q) = t \;\wedge\; \exists\, q \in Q \;\; \mathrm{valid}(q) \}, \]
and for any $Q, Q' \in \xs'$
\[
\begin{array}{r@{\;}l@{\;}lll@{\;}l}
	Q & [\opsem]'^a & Q' & \iff \exists\, t \in T_\Sigma	& Q' = \{ q' : q \opsem^a q', 	& q \in Q, \cterm(q') = t \} \\
\end{array}
\]
\end{definition}

In summary, the states of $\ltmsl'_\alpha$ are sets of execution states with a common projection, and the successors of these sets are the union of the successors of their elements grouped by their subject terms and by the action.

\begin{theorem} \label{thm:malpha}
	$\ltmsl'_\alpha$ and  $\mathcal U(\mathcal M, \lambda_{E(\alpha)})$ are bisimilar Kripke structures.
\end{theorem}

\Cref{thm:malpha} tells that $\ltmsl'_\alpha$ is an effective candidate to check branching-time properties on Maude specifications with strategies according to the ideas of~\cref{sec:idea}. All these structures and propositions have been stated in terms of labeled transition systems, while state-based logics like LTL, CTL, and CTL* are defined on unlabeled transition systems. As we mentioned in~\cref{sec:strategies}, this is without loss of generality, because unlabeled transition systems can be viewed as labeled ones with a single arbitrary label. However, it is important that we forget about the labels of a labeled transition system before checking state-based properties, not only by efficiency reasons, but also by semantic ones. Otherwise, the labels will change the model semantics as the strategy continuations did in the previous section. For instance, suppose a strategy \texttt{r1 ; r2 | r3 ; r4} is applied to a term $t_1$ with the rules $t_1 \to^{\texttt{r1}} t_2$, $t_1 \to^{\texttt{r3}} t_2$, $t_2 \to^{\texttt{r2}} t_3$, and $t_2 \to^{\texttt{r3}} t_4$. The CTL property $\ctlAll \ctlNext \ctlOne \ctlEvly t_4$ will not be true if edge labels are considered, but it will if they are not, as it should be for state-based logics. On the contrary, for logics that operate on labeled transition systems like $\mu$-calculus, the edge labels should be preserved and used to distinguish successor states when they are merged, because our notion of strategy $\lambda : (S \cup A)^+ \to \mathcal P(A \times S)$ conditions the next steps on the previous actions too. Using the same example, $\mucOne{\texttt{r1}} \mucOne{\texttt{r4}} \top$ should only be true if \texttt{r4} can be applied after \texttt{r1}. With these precautions, the following corollary claims that we can check CTL, CTL*, and $\mu$-calculus properties, among others, using $\ltmsl'_\alpha$.

\begin{corollary} \label{cor:satism}
	$(\mathcal M, E(\alpha, t)) \vDash \varphi \iff \ltmsl'_\alpha \vDash \varphi$ for any bisimilarity-invariant temporal property $\varphi$.
\end{corollary}

	The generated transition system $\ltmsl_\alpha$ is finite and its transition decidable if the reachable states from the initial one are finitely many~\cite{fscd}. Since merged states are the combinations of normal execution states, the number of states can grow exponentially at worst, although it would usually decrease, like in the vending machine example. 

\begin{corollary} \label{cor:decidable}
	If the reachable states from $t \ao \alpha$ by $\to_{s,c}$ are finitely many, $(\mathcal M, E(\alpha, t)) \vDash \varphi$ is decidable for LTL, CTL*, and $\mu$-calculus.
\end{corollary}
 \section{Model checking using external model checkers} \label{sec:maudesmc}

The extension of the Maude LTL model checker for strategy-controlled specifications~\cite{fscd} generates as part of its job a labeled transition system, the $\ltmsl_\alpha$ of~\cref{sec:example}. With the adaptations described in~\cref{sec:vending}, this LTS can be transformed into $\ltmsl'_\alpha$, where branching-time properties can be properly checked. Thanks to the modular design of the original model checker, adopted by our extension, this model is exposed as an abstract Kripke structure where the successors and the atomic properties satisfied by a state can be queried using C++ functions. Hence, model checking properties in other logics only requires implementing their algorithms and the adaptations on top of this interface. However, instead of writing our own model-checking algorithms, we have found convenient to reuse already used and tested implementations for the target logics, since they are ultimately based on Kripke structures. A good candidate is the language-independent model checker LTSmin~\cite{LTSmin}, which is able to efficiently interact with our Kripke-like representation of the model on the fly at the C++ level and supports all logics we have considered here, CTL* and $\mu$-calculus. In addition, we have established connections with other model checkers like NuSMV~\cite{NuSMV}, the \texttt{pyModelChecking}~\cite{pymodelchecking} library, and Spot~\cite{spotomega}, and we have also written our own implementation of a $\mu$-calculus algorithm. More details about these connections are given at the end of this section. Additional logics and backends can be added without much effort using this approach.

	Aiming at discharging users from learning the particular syntax and mode of operation of the different backends, a common and simplified interface is provided by the \emph{unified Maude model-checking tool} \texttt{umaudemc}~\cite{umaudemc}. This program has a graphical and a command-line interface where the model-checking problem data is entered and the results are shown. The command for checking a property is the following:
\begin{quotation}
\newcommand*\nonterm[1]{$\langle\,$\textit{#1}$\,\rangle$}
\newcommand*\terminal[1]{\texttt{#1}}
\leavevmode\kern-1.8em\terminal{umaudemc} \terminal{check} \nonterm{file name} \nonterm{initial term} \nonterm{formula} [ \nonterm{strategy} ]
\end{quotation}
The formula can be expressed in a syntax that extends the predefined Maude \texttt{LTL} module with operators for CTL* and $\mu$-calculus. In the first case, the only new constructors are the universal \texttt{A\_} and existential \texttt{E\_} path quantifiers. For the $\mu$-calculus, the syntax is extended with the universal modalities \texttt{[\_]\_} and \texttt{[.]\_}, the existential modalities \texttt{<\_>\_} and \texttt{<.>\_}, the fixed-point operators \texttt{mu\_.\_} and \texttt{nu\_.\_}, and variables.
\begin{lstlisting}
*** CTL and CTL*
op A_ : Formula -> Formula [ctor prec 53] .
op E_ : Formula -> Formula [ctor prec 53] .

*** mu-calculus
subsort @MCVariable@ < Formula .

op <.>_  : Formula -> Formula [ctor prec 53 format (c o d)] .
op [.]_  : Formula -> Formula [ctor prec 53 format (c d d os d)] .
op <_>_  : @ActionSpec@ Formula -> Formula [ctor prec 53 ...] .
op [_]_  : @ActionSpec@ Formula -> Formula [ctor prec 53 ...] .
op mu_._ : @MCVariable@ Formula -> Formula [ctor prec 64] .
op nu_._ : @MCVariable@ Formula -> Formula [ctor prec 64] .

*** Action lists
sorts @ActionSpec@ @ActionList@ .
subsort @ActionList@ < @ActionSpec@ .
op __ : @ActionList@ @ActionList@ -> @ActionList@ [ctor assoc] .
op ~_ : @ActionList@ -> @ActionSpec@ [ctor] .
\end{lstlisting}
Modalities are generalized so that they can take one or more rule labels of the module as actions, separated by space. This follows the widespread notation $\mucAll{C} \varphi \coloneq \bigwedge_{a \in C} \, \mucAll{a} \varphi$ and $\mucOne{C} \varphi \coloneq \bigvee_{a \in C} \, \mucOne{a} \varphi$. In case $C$ is the complete set of actions, a dot can be written instead.
The complement of the list of actions can be specified by preceding it with the negation symbol \verb|~|. Variables for $\mu$-calculus can be any token that does not conflict with the other elements in the formula. The sorts \texttt{@ActionList@} and \texttt{@MCVariable@} are populated at the metalevel before parsing, based on the rule labels of the target module and on a previous scan of the formula. The \texttt{umaudemc} tool parses the input formula within this Maude signature, deduces the least-general logic this formula belongs to, and then calls the appropriate backend with the appropriate configuration.

To illustrate its usage, we will check some branching-time properties of the river-crossing puzzle. The CTL formula $\mathbf{A} \, \ctlAllw \mathbf{E} \, \ctlEvly goal$ expresses that every state of the river-crossing puzzle can be continued to a solution. This formula is satisfied when the system is controlled by the \texttt{safe} strategy, but not when using the \texttt{eagerEating} strategy or when the system runs uncontrolled.

\begin{lstlisting}[language={}]
$ umaudemc check river.maude initial 'A [] E <> goal' safe
The property is satisfied in the initial state
(16 system states, 264 rewrites).

$ umaudemc check river.maude initial 'A [] E <> goal' eagerEating
The property is not satisfied in the initial state
(43 system states, 4012 rewrites).

$ umaudemc check river.maude initial 'A [] E <> goal'
The property is not satisfied in the initial state
(36 system states, 3058 rewrites).
\end{lstlisting}
The reason is that no solution can be reached once a character has been eaten, which may happen in the last two cases. Counterexamples are only shown if the selected backend supports them, and the \texttt{-c} flag can be used to prefer one of these. The following counterexample confirms our explanation for the refutation of the last property.\footnote{For a branching-time logic, counterexamples can be provided for purely universal formulae and examples for purely existential formulae. In case both quantifications are mixed, a prefix of the path until the second quantifier applies can be given.}
\begin{lstlisting}[language={}, mathescape, escapechar=^]
^\$^ umaudemc check river.maude initial 'A [] E <> goal' -c
The property is not satisfied in the initial state
(36 system states, 125 rewrites)
| right | left shepherd wolf goat cabbage
$\vee$ ^\itshape rl G' | shepherd G => G | shepherd G' [label alone] .^
| right shepherd | left wolf goat cabbage
$\vee$ ^\itshape rl shepherd G' | wolf goat G => shepherd G' | wolf G [label wolf-eats]^ .
O right shepherd | left wolf cabbage
\end{lstlisting}
However, the property $\ctlAll \ctlAllw (\mathit{risky} \,\vee\, \mathit{death} \,\vee\, \ctlOne \ctlEvly \mathit{goal})$ holds under the \texttt{eagerEating} strategy.
\begin{lstlisting}[language={}]
$ umaudemc check river.maude initial \
    'A [] (risky \/ death \/ E <> goal)' eagerEating
The property is satisfied in the initial state
(43 system states, 1088 rewrites).
\end{lstlisting}

We can also check $\mu$-calculus properties, like the fact that the only initial movement not leading to a risky state is \texttt{goat}:
\begin{lstlisting}[language={}]
$ umaudemc check river.maude initial \
    '[ alone wolf cabbage ] risky /\ < goat > ~ risky'
The property is satisfied in the initial state
(5 system states, 18 rewrites, 15 game states).
\end{lstlisting}
Then, we wonder if the goal can be reached without moving the goat again: this is the property $[ \mathtt{goat} ] \, ( \mu Z . \, goal \,\vee\, \langle\mathtt{alone} \; \mathtt{wolf} \; \mathtt{cabbage} \rangle \, Z)$ where the fixed-point subformula describes the states where the goal can be reached using any sequence of moves other that \texttt{goat}. The answer is no  if the rules of the game are respected as in the \texttt{eagerEating} strategy:
\begin{lstlisting}[language={}]
$ umaudemc check river.maude initial \
    '[ goat ] (mu Z . goal \/ < ~ goat > Z)' eagerEating
The property is not satisfied in the initial state
(43 system states, 192 rewrites, 364 game states).
\end{lstlisting}
Notice that we have replaced the list of labels \texttt{alone wolf cabbage} by \verb|~ goat| to illustrate the complement notation for actions. These are not exactly the same, because the complement of \texttt{goat} also includes the rules \texttt{wolf-eats} and \texttt{goat-eats}, but they do not change the satisfaction of the property.
On the contrary, the uncontrolled system satisfies the formula, since it can pass by forbidden states:
\begin{lstlisting}[language={}]
$ umaudemc check river.maude initial \
    '[ goat ] (mu Z . goal \/ < ~ goat > Z)'
The property is satisfied in the initial state
(33 system states, 168 rewrites, 362 game states).
\end{lstlisting}

While the \texttt{umaudemc} tool automatically enables the branching-time adaptations of the model according to the input formula, these defaults can be overwritten with the \texttt{--purge-fails} and \texttt{--merge-states} options. Coming back to the vending machine example of~\cref{sec:vending}, with the \texttt{merge-states} adaptation disabled, we can see that the CTL property $\ctlAll \ctlNext \ctlOne \ctlEvly \mathit{hasCake}$ is not satisfied when the system is controlled by the strategy $\alpha$, but it is when controlled by the equivalent strategy $\beta$:
\begin{lstlisting}[language={}]
$ umaudemc check vending.maude initial 'A O E <> hasCake' \
	'put1 ; apple | put1 ; put1 ; cake' --merge-states=no
The property is not satisfied in the initial state
(6 system states, 72 rewrites)

$ umaudemc check vending.maude initial 'A O E <> hasCake' \
	'put1 ; (apple | put1 ; cake)' --merge-states=no
The property is satisfied in the initial state
(5 system states, 60 rewrites).
\end{lstlisting}
However, when states are properly merged, the property is satisfied for both strategy expressions as follows from~\cref{cor:satism}:
\begin{lstlisting}[language={}]
$ umaudemc check vending.maude initial 'A O E <> hasCake' \
	'put1 ; apple | put1 ; put1 ; cake'
The property is satisfied in the initial state 
(6 system states, 52 rewrites).
\end{lstlisting}
The transition systems generated for each model with the different adaptations can be observed with the \texttt{umaudemc graph} command. For instance, \texttt{umaudemc graph river.maude initial safe} would generate something similar to \cref{fig:safe}, and its states in blue can be removed with the \texttt{--purge-fails=yes} option.

\subsection{The architecture of \texttt{umaudemc}}

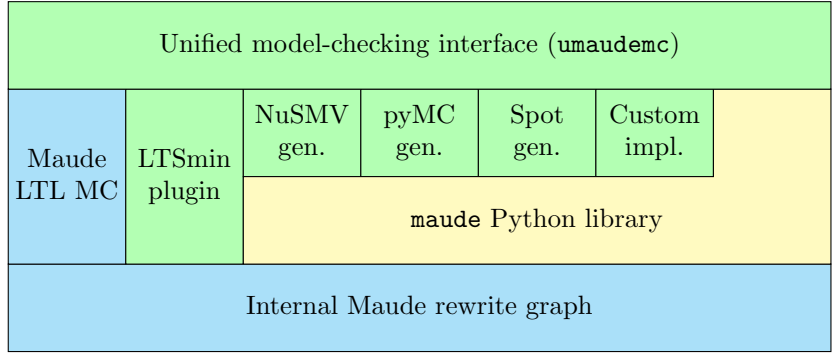
\begin{figure}
\centering
\begin{tikzpicture}[x=4.4em,y=3.3em]
\draw[fill=cyan!30] (0, 0) rectangle +(7, 1);
\node[anchor=center] at (3.5, 0.5) {Internal Maude rewrite graph};
\draw[fill=cyan!30] (0, 1) rectangle +(1, 2);
\node[anchor=center] at (0.5, 2) {\begin{tabular}c Maude \\ LTL MC\end{tabular}};
\draw[fill=green!30] (1, 1) rectangle +(1, 2);
\node[anchor=center] at (1.5, 2) {\begin{tabular}c LTSmin \\ plugin\end{tabular}};
\draw[fill=green!30] (2, 2) rectangle +(1, 1);
\node[anchor=center] at (2.5, 2.5) {\begin{tabular}c NuSMV \\ gen.\end{tabular}};
\draw[fill=green!30] (3, 2) rectangle +(1, 1);
\node[anchor=center] at (3.5, 2.5) {\begin{tabular}c pyMC \\ gen.\end{tabular}};
\draw[fill=green!30] (4, 2) rectangle +(1, 1);
\node[anchor=center] at (4.5, 2.5) {\begin{tabular}c Spot \\ gen.\end{tabular}};
\draw[fill=green!30] (5, 2) rectangle +(1, 1);
\node[anchor=center] at (5.5, 2.5) {\begin{tabular}c Custom \\ impl.\end{tabular}};
\draw[fill=yellow!30] (2, 1) -- (7, 1) -- (7, 3) -- (6, 3) -- (6, 2) -- (2, 2) -- (2, 1);
\node[anchor=center] at (4.5, 1.5) {\texttt{maude} Python library};
\draw[fill=green!30] (0, 3) rectangle +(7, 1);
\node[anchor=center] at (3.5, 3.5) {Unified model-checking interface (\texttt{umaudemc})};
\end{tikzpicture}
\caption{Architecture of the \texttt{umaudemc} model-checking tool.} \label{fig:arch}
\end{figure}

	As we have seen with the examples of the previous section, the \texttt{umaudemc} tool allows checking temporal properties on both standard and strategy-controlled Maude specifications regardless of which model-checking backend is doing the job behind the scenes. All of them rely on the internal Maude rewrite graph  used by the Maude LTL model checker~\cite{maudemc} and by our extension for strategy controlled systems~\cite{fscd}, which correspond to the C++ classes \texttt{StateTransitionGraph} and \texttt{StrategyTransitionGraph} in their implementations. As illustrated in~\cref{fig:arch}, some backends access these graphs directly while others use a Python library called \texttt{maude}~\cite{maude-bindings} that we have developed for this and other projects. This library exposes all relevant Maude entities and operations as objects and methods in Python by directly interacting with the Maude implementation at the binary level,\footnote{The \texttt{maude} library is a language binding implemented using the SWIG interface generator. More details are available in its repository~\cite{maude-bindings}.} including the strategy-controlled and the standard rewrite graphs. Exploring these graphs and evaluating atomic propositions on them, models are generated for the various supported backends.

	The \texttt{maude} library is also used directly by the \texttt{umaudemc} tool to process the problem data, the verification results and the counterexamples, and to produce printable graphs of the models. The extended language of temporal properties admitted by the tool is specified in a Maude module, parsed using the library, and translated to the syntax of the temporal properties supported by the selected backend. Whether the adaptations of~\cref{sec:vending} are applied or not is also decided depending on the problem data, and they are implemented in C++ inside the LTSmin plugin or in Python for the backends based on the \texttt{maude} library.	The \texttt{umaudemc} tool will detect which backends are installed and call the most convenient for each supported logic, although the search order can be changed with the \texttt{--backend} option. In addition to the Maude LTL model checker and LTSmin, which is described in~\cref{sec:ltsmin}, the available backends are:
\begin{itemize}
	\item NuSMV~\cite{NuSMV}, which supports LTL and CTL. The model is communicated by writing a low-level specification file in the NuSMV format. It calculates counterexamples for CTL properties too.
	\item \texttt{pyModelChecking}~\cite{pymodelchecking} is a Python library that targets LTL (by the tableau method), CTL, and CTL* model checking. The Kripke structure is constructed as a Python object from the Maude model.
	\item Spot~\cite{spotomega} is a C++ framework for LTL and $\omega$-automata manipulation with a Python library. Models are built as Kripke structures using this library, but it admits more complex $\omega$-automata. It also admits on-the-fly model checking, but not through the Python interface.
	\item Our own implementation in Python of the $\mu$-calculus model-checking algorithm in~\cite{mucalcmc}, using the Zielonka algorithm~\cite{zielonka} for parity game solving.
\end{itemize}
\Cref{fig:logics} summarizes which logics can be checked with each backend. Although LTSmin supports all logics we have considered, other model checkers are easier to install, provide more informative output, or exhibit better performance in some cases despite their less efficient connection, as discussed in~\cref{sec:evaluation}. Adding connections to other model checkers and logics is relatively simple, as suggested by the number of code lines written for each backend in~\cref{fig:logics}, since the models described in~\cref{sec:example,sec:vending} are easily accessible and compatible in principle with any logic.

\begin{table}\centering
\begin{tabular}{rccccc} \toprule
			& LTL 		& CTL 	& CTL*	& $\mu$-calculus	& Lines \\ \midrule
	Extended Maude	& on-the-fly	&	&	&		& 1200	\\
	LTSmin		& on-the-fly	& X	& X	& X		& 1140	\\
	pyModelChecking	& tableau	& X	& X	&		& 147	\\
	NuSMV		& tableau	& X	&	&		& 199	\\
	Spot		& automata	& 	&	&		& 203	\\
	Builtin		& 		& X	&	& X		& 400	\\ \bottomrule
\end{tabular}
\caption{Logics supported by the backends in \texttt{umaudemc}.} \label{fig:logics}
\end{table}

\subsection{The LTSmin language plugin} \label{sec:ltsmin}

	LTSmin~\cite{LTSmin} is a collection of generic model-checking programs that can operate on models expressed in different specification languages. These models are exposed as Kripke structures by some builtin or pluggable language modules using its \emph{Partitioned Next State Interface} (PINS). In order to check properties with this toolset, we have implemented a language module for Maude. The module \texttt{libmaudemc} is a shared C library linked with the implementation of Maude\footnote{Maude is usually distributed as a single binary, but we have built it as a shared library \texttt{libmaude} to distribute the interpreter and this plugin together without including twice the same executable code.} that exports the functions required by the PINS interface. Model checking a temporal property using LTSmin and the Maude plugin consists of the following steps:
\begin{enumerate}
	\item The \texttt{pins2lts-*} model-checking tools of LTSmin are called with the problem data and with a \texttt{--loader} argument indicating the path of the Maude plugin. The Maude language module is loaded in memory using the POSIX's \texttt{dlopen} API, so that its exported functions and global variables required by the PINS interface can be accessed. One of these functions is called to pass the Maude-specific problem data to the plugin (the initial term, the strategy, and some other parameters) and prepare the Kripke structure that will be made available to the model-checking algorithms.

	\item When the model-checking algorithm for the given logic wants to know if an atomic property $p$ is satisfied in a state, it calls the \texttt{state\_label} function of the plugin that evaluates the term \lstinline[mathescape,basicstyle={\ttfamily}]{$\cterm(Q)$ |= $p$} and returns its Boolean result. When the model checker requires the successors of a state, it calls the \texttt{next\_state} function that enumerates them with their corresponding edge labels.

	\item The verification result is printed to the terminal. In the $\mu$-calculus case, a parity game is generated instead, which has to be solved by an external tool from the mCRL2 project~\cite{mCRL2}.
\end{enumerate}
The Kripke structures presented to LTSmin are $\mathcal M$, $\mathcal M_\alpha$ or $\mathcal M'_\alpha$ depending on whether the model is controlled by a strategy or not, and on the arguments \texttt{--purge-fails} and \texttt{--merge-states} passed to the language plugin. In fact, the \texttt{next\_state} function called in the second step is chosen at the beginning from a small set of alternative functions that implement the adaptations described in~\cref{sec:vending} for state-based or edge-based branching-time logics. This choice could be inferred from the temporal formulae to be checked, but this information is never passed to the language plugin. The same happens with the atomic propositions, which must be supplied directly with the \texttt{--aprops} argument, since the set of potential atomic proposition can be infinite as they are regular Maude operators with parameters. These disadvantages and the choice of an appropriate LTSmin command for a given property are avoided when using the \texttt{umaudemc} utility. Among the different programs included in LTSmin, \texttt{umaudemc} invokes its sequential explicit-state model checker \texttt{pins2lts-seq} for LTL and $\mu$-calculus properties with action specifications,\footnote{Notice that $\mu$-calculus properties that refer to both edge and state labels cannot be verified with the last LTSmin version at the moment of writing (3.0.2). We have proposed a change that makes it possible and the modified version is available for download in~\cite{stratweb} in the meantime.} and the symbolic model checker \texttt{pins2lts-sym} for CTL, CTL*, and $\mu$-calculus properties without action specifications (i.e.\ using only the \texttt{<.>} and \texttt{[.]} modalities).

	The Maude language module does not take full advantage of LTSmin. Its PINS interface allows representing states as vectors of integer indices and declaring dependencies between their entries, so that the model-checking algorithms can use them for better efficiency and parallelization. However, our plugin's states are single indices to the internal Maude rewrite graph. Automatically partitioning an arbitrary Maude specification and inferring dependencies between the resulting parts seems to be a very complex task.

\section{Evaluation} \label{sec:evaluation}

We have tested and compared the performance of the model-checking backends described in this article using the collection of strategy-controlled Maude specifications and temporal properties available at the Maude strategy language website~\cite{stratweb}. Most of these examples are relatively small (classical concurrency problems, games, models translated from other model-checking tools, etc) and they have been specifically written to test our model checker, but some others have a greater size and interest on their own. The results cannot be interpreted as a comparison of the model-checking tools themselves, since the figures also reflect the efficiency of the connections to our Maude models. The reader should also keep in mind from~\cref{sec:maudesmc} that they all operate on the same Kripke structure produced by Maude from a strategy-controlled specification, with some common adjustements in the case of branching-time properties. In the same website, the complete listing of the test cases and their results are available for reproducibility, and they can be executed with the \texttt{test} subcommand of the \texttt{umaudemc} tool.

\begin{figure}
	\leavevmode\kern-3em
	\hbox{\includegraphics[page=1, scale=0.5, clip, trim={0 0 1cm 0}]{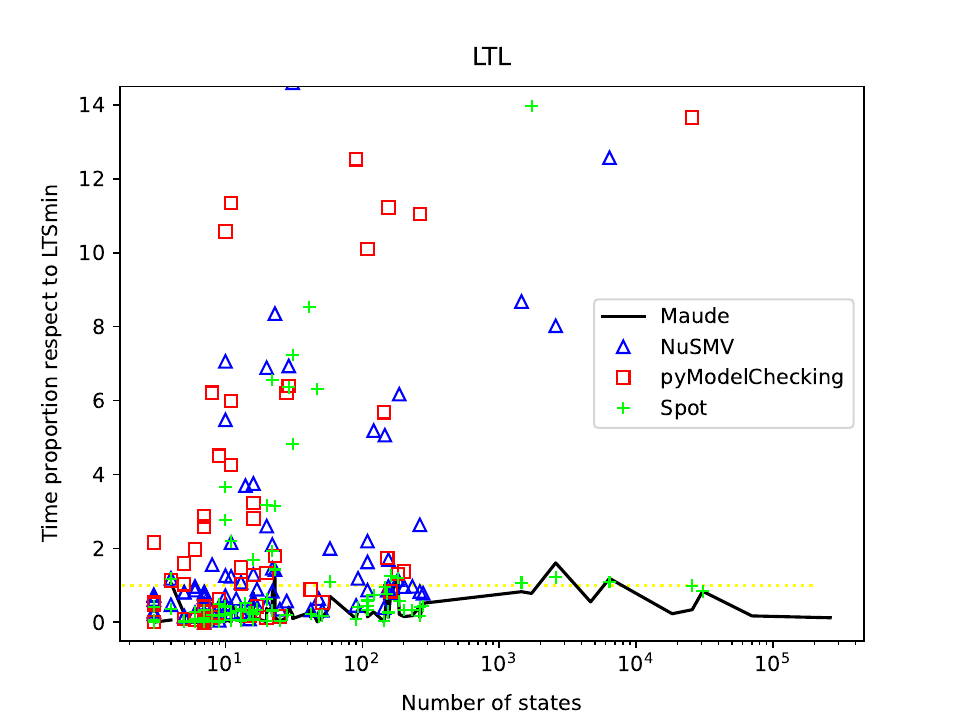}
	\includegraphics[page=2, scale=0.5]{img/perfplot.pdf}}
	\caption{Compared performance of the model-checking backends.} \label{fig:perfbackends}
\end{figure}

	The plots in~\cref{fig:perfbackends} compare the time spent by the different backends to execute the same model-checking problems ordered by their number of states. These problems are given by a Maude module, an initial term, a strategy expression, and a temporal formula. For every test case $c$ and backend $b$, the plot shows a specific marker determined by $b$ in the coordinates
\[ (n^\text{LTSmin}_c, (t^b_c - t^b_e) \;/\; (t^\text{LTSmin}_c - t^\text{LTSmin}_e)), \]
where $n^b_c$ and $t^b_c$ are respectively the number of states and the execution time of the test case $c$ in the backend $b$, and $e$ is an empty test case with a single state and a trivial property. In other words, looking at a fixed vertical rule we can compare the performance of the different backends for a test case, since the height of the marks indicates the proportion of time a backend has taken to complete its task respect to LTSmin, so that higher means worse. LTSmin has been chosen as a common reference since it supports all considered logics, and so it can run all test cases. Even though the number of states are referred to a fixed backend, this figure is essentially a property of the test case and it is usually the same for all tools.\footnote{The number of states may differ in test cases where the temporal property does not hold, since counterexamples can be found sooner or later by the different on-the-fly implementations, and in some corner cases explained in the following.}
Moreover, we have subtracted the initialization time $t^b_e$ before calculating the coefficients, because small examples are highly influenced by the quite different initialization times of the backends, with LTSmin being a thousand times slower on the empty example $e$ than the Maude LTL model checker. In order to compare the new supported model checkers with the builtin Maude one, the left plot includes tests against linear-time properties. The results on our smaller collection of branching-time properties are shown in the right plot.

	For LTL properties, the Maude model checker is usually and expectedly the fastest, since it is directly connected to the rewrite graphs. However, Spot is often and LTSmin sometimes very close.
The peaks in the Maude curve above $10^3$ states are caused by the different order in which states are explored by the backends. Although all these cases evaluate properties that are satisfied, in which the whole state space has to be expanded, some exploration orders may detect the equivalence of two states earlier in some corner cases.
Surprinsingly, the Python-based algorithms are more efficient than LTSmin, but only with small examples. NuSMV and \texttt{pyModelChecking} do not behave bad for lower sizes, even though their algorithms do not operate on the fly, but they do not terminate in reasonable time and memory limits when the problems are big enough.

	Regarding branching-time properties, although the size of the examples is small, we observe that LTSmin exhibits the worst performance and all other backends check the same test cases in half of the time. However, for the biggest $\mu$-calculus problem checked, its performance is better than that of our builtin backend.

\section{Related work}  \label{sec:relatedwork}

	\newcommand*\cparagraph[1]{\medskip\noindent\textbf{#1.}}

	Three independent but related topics are addressed in this work, model checking, strategies and rewriting, which have had fruitful interactions. In this section, we review on related work the combinations of these topics towards approaches that are close to what has been presented here.

	\cparagraph{Strategies and rewriting} Strategies are inherent to rewriting and reduction, and so their study dates back to the origins of $\lambda$-calculus. When specifying the behavior of algorithms and other systems, the Kowalski's motto \emph{Algorithm = Logic + Control}~\cite{kowalski} is translated in this context to the \emph{Rule + Strategies} approach~\cite{pettorossi,lescanneOrme}, where strategies express an additional level of specification that controls the rule rewriting system compositionally and without mixing their concerns. In addition to Maude's, other strategy languages have appeared like ELAN~\cite{elan}, Stratego~\cite{stratego}, TOM~\cite{tom}, and $\rho$Log~\cite{rholog} for term rewriting, and Porgy~\cite{porgyJournal} for graph rewriting. Strategy-controlled specifications have been used to describe many examples of systems from different fields~\cite{eden,sudoku,completion,pssm,chemicalStrat,srewSocialNetworks}. However, the verification techniques used for these tools do not include model checking as understood here.

	\cparagraph{Model checking and rewriting} Various model checkers have been proposed for rewriting systems in the Maude context. The main one is the Maude LTL model checker~\cite{maudemc} integrated in the Maude interpreter and applied on many real models, among others~\cite{cassandra,ikp,coreerlang}. In addition, $\mu$-calculus model checkers have been once implemented in Maude itself~\cite{wangMucalc,mumaude}, the Real Time Maude~\cite{realTimeMaude} framework includes a Timed CTL model checker for real-time systems, the Maude LTL logical model checker~\cite{maudeLogicalMC} symbolically verifies infinite-state systems using narrowing, and some fragments of the Temporal Logic of Rewriting (TLR*)~\cite{tlr}, whose relation with strategies is discussed below, can be checked with different implementations~\cite{maudeLTLRFair,oscarTLR}. Model checking has also been used in the CafeOBJ~\cite{cafeobj,cafeobj-smv} language, despite not including a dedicated model checker.

	\cparagraph{Model checking and strategies} The relation between model checking and strategies is a wide and active research topic in the context of games, multiagent and open systems, where strategies are usually followed by the players or agents to achieve some defined goal and where many properties can be expressed in terms of strategies. 
Representative logics are the \emph{Alternating-Time Logics} ATL and ATL*~\cite{atl} that respectively include the usual CTL and CTL* operators, but whose path quantifiers $\ctlOne$ and $\ctlAll$ are replaced by the strategic modalities $\lAngle A \rAngle$ and $\lBrack A \rBrack$. The meaning of $\lAngle A \rAngle \varphi$ is that a strategy can be chosen for each of the agents in the coalition $A$ to make $\varphi$ hold, regardless of what the agents not in $A$ do. The more expressive \emph{Strategy Logic} (\textsf{SL})~\cite{mogaveroJournal} of Mogavero, Murano, Perelli, and Vardi is extended with strategy variables that can be existentially $\lAngle x \rAngle \varphi$ and universally $\lBrack x \rBrack \varphi$ quantified and then assigned $(a, x) \, \varphi$ to one or more agents $a$. The satisfaction of their formulae is defined recursively, and so they have a concept of strategy-controlled checking to evaluate the $\varphi$ and $(a, x) \, \varphi$ subformulae that coincides with our~\cref{def:btime}, or more precisely, to the generalized CTL* semantics of~\cref{sec:genlogics}. However, the semantics of these logics is crucially influenced by the type of strategy considered~\cite{moduleCheckingStrats,strategyVariants}, which in most cases is deterministic and sometimes memoryless, although the literature on variations of strategy logics is extensive, and the case of general intensional strategies are covered in logics like USL~\cite{revocableRefinable} and \textsf{SL}$^\prec$~\cite{nondetermSl}. Our model-checking problem cannot be directly seen as a particular case of the problem for these logics, because strategies do not appear explicitly and they are always bound by quantifiers. On the contrary, the associated satisfaction problem for CTL* properties, whether there is a strategy such that the formula is satisfied under its control, can be expressed as a very particular case of the last two mentioned logics. In their full generality, model checkers have not been implemented for these logics as far as we know since their problems are very hard, but some subsets can be effectively model checked. MCMAS~\cite{mcmas} is an extensive open-source model checker for multi-agent systems that supports ATL as verification logic, and restrictions of Strategy Logic through extensions~\cite{mcmas-sl1g}. In addition to the verification result, they can also synthesize strategies for the agents that make the formula hold. Strategy synthesis is related to the so-called \emph{controller synthesis}~\cite{controllerSynthesis} with important industrial applications.
In a similar but different context lies \textsc{Uppaal Stratego}~\cite{uppaalStratego} from the \textsc{Uppaal}~\cite{uppaalmc} modeling environment for real-time systems. This tool allows synthesizing strategies to make a property hold as in the other mentioned tools, but these strategies can be later used to execute the constrained system and model check it against other properties, considering only this restricted \emph{strategy-space}. Strategies in this case are memoryless and deterministic, but they follow the same idea of this work.

	Strategies have also been applied to reduce the search space for the sole purpose of model checking, by guiding its search to a counterexample or witness of the desired property, like search heuristics~\cite{heuristics}. For example, this has been done~\cite{reachabilityExprs} with a language of reachability expressions including union, concatenation and iteration operators that resemble those of Maude and similar strategy languages.

	\cparagraph{Temporal Logic of Rewriting and other logics} Meseguer's Temporal Logic of Rewriting (TLR*)~\cite{tlr} is also connected with strategies and the Maude strategy language. This logic extends CTL* with \emph{spatial action patterns} that symbolically designate a collection of rule applications. They can be used in path formulae to indicate how the next transition to be executed should be, and so TLR* is at the same time a state-based and edge-based temporal logic. For example, the property $\ctlAllw (\neg \mathrm{goat} \rightarrow \ctlNext \mathrm{risky})$ says that any action other than \texttt{goat} would lead to a risky state. Spatial action patterns can be more complex and include restrictions on the variables and the context where rules are applied. In fact, they are very similar to a combination of rule applications, the \texttt{top} modifier, and the \skywd{matchrew} of the Maude strategy language, to which they can be translated. The relation with strategies comes from the possibility of checking certain properties on infinite-state systems by a strategy-controlled exploration. Guarantee formulae (those only containing the temporal operators $\mdlgwhtcircle$, $\mdlgwhtsquare$, and $\mathbf{U}$ without negations) can be translated to strategy expressions whose evaluation is a semidecision procedure for the original formulae. In Sections 6 and 7 of~\cite{tlr-full}, a strategy language similar to that of Maude is introduced for this particular purpose.
\begin{align*}
		b	&\Coloneq \top \mid \bot \mid p \mid \neg b \mid b \wedge b \mid b \vee b \\
		e	&\Coloneq \mathit{idle} \mid \delta \mid \mathit{any} \mid e \wedge e \mid (e | e) \mid e \, ; e \mid e+ \mid e \ctlUntil e  \mid e.b
	\end{align*}
However, some combinators of this language are neither available nor expressible in the current Maude strategy language, and so it cannot be used to implement these procedures. Notice that the strategy-aware model checker is not needed for that, but only the execution engine of strategies. On the other hand, writing a TLR* model checker for finite-state systems would be reasonably simple using the tools and connections developed in this work.

	Finally, \emph{propositional dynamic logic} (PDL)~\cite{PDL}, \emph{linear dynamic logic} (LDL)~\cite{LDL} and their variations are partially related to the model-checking problem of strategy-controlled systems. Their formulae $\langle r \rangle \, \varphi$ include complex actions $r$ built using edge labels, regular expression combinators, and tests, which can be seen as a subset of the strategy language too. Atomic propositions and more complex temporal formulae can be checked at the end of those sequences of actions or at arbitrary points during them using tests. Unlike in our approach, properties are not checked in the system restricted by the action patterns but at some execution points indicated by them.

\section{Conclusions and future work}

	Strategies are a useful resource for elaborating modular rewriting-based specifications, where simpler rules represent the local transformations of the model, and strategies describe at a higher level restrictions that capture its global behavior, guide them towards a goal, apply them more efficiently, etc. The current version of the Maude specification language~\cite{maude} includes an LTL model checker for rewriting-based specifications and a strategy language to control rewriting, but these are independent and properties cannot be checked on strategy-controlled models. Therefore, in a previous work, we extended the Maude LTL model checker to handle strategy-controlled systems~\cite{fscd}. The fundamental idea is that properties should only be checked on the executions allowed by the strategy, and a small-step operational semantics was defined to determine which are those for an expression in the Maude strategy language. Using this semantics, the original Kripke structure can be transformed to another one whose executions are exactly those allowed by the strategy, in which properties can be checked using standard algorithms. In this paper, we extend the discussion to branching-time properties, realizing that strategies can also be seen as subtrees of the execution trees of the original model where properties can be checked. However, the model transformation proposed for linear-time properties must be adapted to maintain the bisimilarity with the unwinding of the original model, so that properties are soundly checked.

	In order to effectively verify branching-time properties, both on strategy-controlled and standard Maude specifications, we have implemented connections with external model checkers, with support for CTL* and $\mu$-calculus. The infrastructure used to connect these model checkers is valuable by itself, and uses a library that allows manipulating and accessing the Maude entities and models from other programming languages~\cite{maude-bindings}. It can be easily used to connect other visualization or verification tools, and to write programs that use Maude as a formal engine. All model checkers can be transparently accessed through a unified \texttt{umaudemc} tool~\cite{umaudemc} that provides extended information and graphical representations of the models and counterexamples. The performance of the connections to the external model checkers is comparable to the builtin Maude model checker.

	This work can be extended with more logics and model-checking backends. Adding some would be specially affordable with the new tools, like the Property Specification Logic~\cite{psl}, already supported by some of the current backends, Meseguer's Temporal Logic of Rewriting and $\mu$-calculus of rewriting. We can also relax the restriction to intensional strategies and try an alternative interpretation of the Maude strategy language, making the iteration behave as a Kleene star, so that fairness constraints can be expressed in the strategy itself. Exploring the satisfaction and strategy synthesis problem mentioned when discussing strategic logics could be another direction of future work.
 
\paragraph{Declaration of competing interest}

	The authors declare that they have no known competing financial interests or personal relationships that could have appeared to influence the work reported in this paper.

\paragraph{Acknowledgements}

	Research partially supported by MCI Spanish projects \emph{TRACES} (TIN2015-67522-C3-3-R) and ProCode-UCM (PID2019-108528RB-C22). Rubén Rubio is partially supported by MU grant FPU17/02319.

\appendix
\section{Proofs}

We do not include all the details in the proofs, which are usually tedious inductive checks, but only sketch the main ideas and the most problematic steps.

\begin{lemma} \label{lema:buchi}
	Every closed $\omega$-language is recognized by a deterministic Büchi automaton with trivial acceptance conditions.
\end{lemma}

\begin{proof}
	Since the language is $\omega$-regular, there must be a Büchi automaton that recognizes it. Since the language is closed, the limits of all executions are allowed, so the acceptance conditions if any are superfluous and can be removed (we have proved this in~\cite{fscd}). The automaton with trivial acceptance conditions can then be determinized by the powerset construction used for finite automata, since the obstacle that impedes determinizing arbitrary Büchi automata are the Büchi conditions.
\end{proof}

\begin{lemma} \label{lemma:bsexec}
	If $\mathcal K$ and $\mathcal K'$ are bisimilar, $\bigcup_{s \in I} \ell(\Gamma_{\mathcal K, s}) = \bigcup_{s \in I'} \ell'(\Gamma_{\mathcal K', s})$.
\end{lemma}

\begin{proof}
	$\mathcal K$ and $\mathcal K'$ are interchangeable in the lemma, so it is enough to prove $G \subseteq G'$, if $G$ and $G'$ are the unions in the statement, by induction with the property $p(ws) = \exists i' \in I', w's' \in \Gamma_{\mathcal K', i'} \quad \ell(ws) = \ell'(w's') \wedge (s, s') \in B$ for all $i \in I$ and $ws \in \Gamma_{\mathcal K, i}$, where $B$ is a bisimulation. The infinite executions are the limits of the finite ones, so they coincide too.
\end{proof}

\setcounter{theorem}{0}
\begin{theorem}
	Given an intensional strategy $\lambda$, there is a finite Kripke structure $\mathcal K'$ bisimilar to $\mathcal U(\mathcal K, \lambda)$ if $E(\lambda)$ is $\omega$-regular. The converse is not true, but in that case $\ell(E(\lambda))$ is $\omega$-regular.
\end{theorem}

\begin{proof}
In summary, the Büchi automaton for $E(\lambda)$ gives the finite Kripke structure $\mathcal K'$, with transition labels somehow clouding the proof.
If $E(\lambda)$ is $\omega$-regular and closed, there is a deterministic automaton $M = (Q, S \cup A, \delta, \iota, Q)$ for it with trivial Büchi conditions by \cref{lema:buchi}. Moreover, all words accepted by this language alternate states in $S$ with actions in $A$. Let $\mathcal K'$ be $(Q \times S, R', I', AP, \ell \circ \pi_2)$ where $I' = \{ (q, s) : q \in \delta(\iota, s), s \in I \}$, $\pi_2$ is the second projection of the pair, and $((q, s), a, (q', s')) \in R' \Longleftrightarrow \exists \, q_m \in Q \quad q_m \in \delta(q, a) \wedge q' \in \delta(q_m, s')$. $\mathcal U(\mathcal K, \lambda)$ is bisimilar to $\mathcal K'$ by the following relation $B = \{(ws, (q, s)) : q \in \hat\delta(\iota, ws) \}$ where $\hat\delta(q, \varepsilon) = \{ q \}$ and $\hat\delta(q, wx) = \bigcup_{q' \in \hat\delta(q, w)} \hat\delta(q', x)$. First, it is clear that states related by $B$ have the same label. Then, if $ws \to wsas'$ and $(ws, (q, s)) \in B$, the state $(q', s')$ where $q' \in \hat\delta(\iota, wsas')$ satisfies $((q, s), a, (q', s')) \in R'$ and $(wsas', (q', s')) \in B$ (the determinism is $M$ is used here). $\hat\delta(\iota, wsas')$ is not empty because $wsas'$ is a prefix of an execution allowed by the strategy, and so recognized by $M$. The other simulation is proven similarly.

	The converse is not true. For an unlabeled counterexample, take $\mathcal K = (\{a, b\}, \{a, b\}^2,$ $\{a\}, \emptyset, \emptyset)$, $\lambda(w) = \{a,b\}$ if $w = a^n$ for $n$ prime and $\{a\}$ otherwise, and $\mathcal K' = (\{c\}, \{(c, c)\},$ $ \{c\}, \emptyset, \emptyset)$. $E(\lambda)$ is not $\omega$-regular and $\mathcal K'$ is bisimilar to $\mathcal U(\mathcal K, \lambda)$ by the only possible total relation. Hence, we only prove that $\ell(E(\lambda))$ is $\omega$-regular. Given $\mathcal K' = (S', R', I', AP, \ell')$, we define the $\omega$-automaton $M = (Q, \mathcal P(AP) \cup A, \delta, \iota, Q)$ where $Q = S' \times \{0,1\} \cup \{\iota\}$ and $\delta(\iota, P) = \{ (s, 1) : \ell(s) = P, s \in I' \}$, $\delta((s, 1), a) = \{ (s', 0) : (s, a, s') \in R' \}$ and $\delta((s, 0), P) = \{ (s, 1) \}$ if $P = \ell(s)$. It is clear that the runs of $M$ are of the form $\iota\, (s_0, 1) (s_1, 0) (s_1, 1) (s_2, 0) (s_2, 1) \cdots$ and they accept words $s_0 a_1 s_1 a_2 s_2 \cdots$ that coincide with the executions of $\mathcal K'$. By \cref{lemma:bsexec}, the projected executions of $\mathcal K'$ coincide with those of $\mathcal U(\mathcal K, \lambda)$ and these are exactly $\ell(E(\lambda))$. Hence, $M$ is a Büchi automaton for the $\omega$-regular language $\ell(E(\lambda))$. Finally, $\ell(E(\lambda))$ being $\omega$-regular is not enough for the existence of a bisimilar finite $\mathcal K'$. The previous counterexample can be refined to show this.

\end{proof}

\setcounter{proposition}{0}
\begin{proposition}
	Given a CTL* formula $\Phi$, $\mathcal K, s \vDash \Phi$ iff $\Gamma^\omega_{\mathcal K, s} \vDash \Phi$.
\end{proposition}

\begin{proof}
	The key fact is that the possible continuations of any finite execution $ws$ for $\Gamma^\omega_{\mathcal K}$ only depend on its final state $s$, since the executions are unrestricted. Hence, $\sfx{\Gamma^\omega_{\mathcal K, s}}{(w s')} = \Gamma^\omega_{\mathcal K, s'}$ for all $ws' \in \Gamma^*_{\mathcal K, s}$.Then, $\mathcal K, s \vDash \Phi$ iff $\Gamma^\omega_{\mathcal K, s} \vDash \Phi$ and  $\mathcal K, s \vDash \phi$ iff $\Gamma^\omega_{\mathcal K, \pi_0}, \pi \vDash \phi$ can be easily proven by induction on the formula. Almost syntactically, $\Gamma^\omega_{\mathcal K, s}$ can be replaced by $s$ and the strategy can be removed in the path relation to obtain the classical definition. The two properties clearly hold in 1 to 7. In 8 and 9, with $E = \Gamma^\omega_{\mathcal K, \pi_0}$, we can observe that $\sfx{E}{\pi_0\pi_1} = \Gamma^\omega_{\mathcal K, \pi_1}$ and $\pi_1 = (\pi^1)_0$, and that $\sfx{E}{\wprefix\pi{n}} = \Gamma^\omega_{\mathcal K, \pi_n}$ with $\pi_n = (\pi^n)_0$. The induction hypothesis can then be applied.
\end{proof}

\begin{lemma} \label{lemma:flat}
	For every $ws_0 \in S^+$ prefix in $E(\lambda)$, $\sfx {E(\lambda)}{ws_0} = \{ \mathrm{flat}(\pi) : \pi \in \Gamma_{\mathcal U(\mathcal K, \lambda), ws_0} \}$ where $\mathrm{flat}((ws_0)(ws_0s_1)(ws_0s_1s_2)\cdots) \coloneq s_0s_1s_2 \cdots$.
\end{lemma}

\begin{proof}
	Executions in $\mathcal U(\mathcal K, \lambda)$ are of the form $(ws_0)(ws_0s_1)(ws_0s_1s_2)\cdots$ where $s_0s_1 \cdots$ is an execution in $E$. For the $\supseteq$ inclusion, take $\Gamma_{\mathcal U(\mathcal K, \lambda), ws_0} \ni \pi = (ws_0)(ws_0s_1)(ws_0s_1s_2) \cdots$, whose $\mathrm{flat}(\pi) = s_0s_1s_2 \cdots$ and $ws_0s_1s_2 \cdots \in E$ since $s_{n+1} \in \lambda(s_n)$. Hence, $\mathrm{flat}(\pi) = s_0s_1\cdots \in \sfx E{ws_0}$ by definition. For the other $\subseteq$ inclusion, $s_0s_1 \cdots \in \sfx E{ws_0}$ implies $ws_0s_1 \cdots \in E$, so $\pi = (ws_0)(ws_0s_1) \cdots \in \Gamma_{\mathcal U(\mathcal K, \lambda), ws_0}$ and $\mathrm{flat}(\pi) = s_0s_1 \cdots$ is in the set.
\end{proof}

\setcounter{proposition}{1}
\begin{proposition}
	Given $(\mathcal K, E(\lambda))$ and a CTL* formula $\varphi$, $\mathcal U(\mathcal K, \lambda) \vDash \varphi$ iff $\mathcal K, E(\lambda) \vDash \varphi$.
\end{proposition}

\begin{proof}
	We follow an inductive proof on the structure of CTL* formulae with the more general property $\mathcal U(\mathcal K, \lambda), w \vDash \varphi$ iff $\mathcal K, \sfx{E(\lambda)}{w} \vDash \varphi$ for all $w \in S^+$. Path formulae need to be handled simultaneously, so the inductive property also includes $\mathcal U(\mathcal K, \lambda), \pi \vDash \varphi$ iff $\mathcal K, \sfx{E(\lambda)}{\pi_0}, \mathrm{flat}(\pi) \vDash \varphi$ (in the lefthand side executions are successions of growing $S^+$ words while in the righthand side they are successions of $S$ states).
To facilitate reading, we will omit the $\mathcal U(\mathcal K, \lambda)$ and $\mathcal K$ prefix when writing the satisfaction relations, and use $E$ for $E(\lambda)$.
\begin{itemize}
	\item ($p$, atomic propositions) By definition, $\mathcal ws \vDash p$ iff $p \in \ell(s)$, and $\sfx{E}{ws} \vDash p$ iff $p \in \ell(s')$ for all $s'w' \in \sfx{E}{ws} = \{ sw'' : wsw'' \in E \}$. Then, $s'$ can only be $s$ and both conditions coincide.
	\item ($\Phi_1 \wedge \Phi_2$) In the standard side, the conjunction is satisfied iff $w \vDash \Phi_i$ for both $i=1,2$. In the strategy side, this happens iff $\sfx{E}{w} \vDash \Phi_i$. By induction hypothesis on both $\Phi_i$ the equivalence holds.
	\item ($\neg \Phi$) The same inductive argument can be used for negation.
	\item ($\mathbf{A}\, \varphi$) This formula is satisfied iff $\pi \vDash \varphi$ for all $\pi \in \Gamma^\omega_{\mathcal U(\mathcal K, \lambda), w}$ in the $\mathcal U(\mathcal K, \lambda)$ side. In the strategy side, this is $\sfx{E}{w}, \rho \vDash \varphi$ for all $\rho \in \sfx{E}{w}$. Using~\cref{lemma:flat}, all these $\rho$ are exactly those $\mathrm{flat}(\pi)$, and applying the induction hypothesis on $\varphi$, both statements are equivalent.
\end{itemize}
Let $\pi$ be $(ws_0)(ws_0s_1)\cdots$, we then target the path satisfaction cases:
\begin{itemize}

	\item ($\ctlNext \varphi$) We should prove that $\pi \vDash \ctlNext \varphi$ is equivalent to $\sfx E{ws_0}, s_0s_1 \cdots \vDash \ctlNext \varphi$. Their definitions translate these to $\pi^1 \vDash \varphi$ and $\sfx{(\sfx{E}{ws_0})}{s_0s_1}, (s_0s_1\cdots)^1 \vDash \varphi$. But they are equivalent by induction hypothesis on $\varphi$, since $\sfx{(\sfx{E}{ws_0})}{s_0s_1} = \sfx{E}{ws_0s_1} = \sfx{E}{\pi_1} = \sfx{E}{(\pi^1)_0}$ and $(s_0s_1 \cdots)^1 = s_1s_2 \cdots = \mathrm{flat}(\pi^1)$.

	\item ($\varphi_1 \,\mathbf U\, \varphi_2$) The formula holds in the standard sense if there is an $n \in \N$ such that $\pi^n \vDash \varphi_2$ and for all $k$ such that $0 \leq k < n$ then $\pi^k \vDash \varphi_1$. In the strategy side, the formula holds if again there is an $n \in \N$ such that $\sfx{(\sfx E{ws_0})}{s_0 \cdots s_n}, s_ns_{n+1}\cdots \vDash \varphi_2$ and $\sfx{(\sfx E{ws_0})}{s_0 \cdots s_k}, s_ks_{k+1}\cdots \vDash \varphi_1$ for all $0 \leq k < n$. Since $\sfx{(\sfx E{ws_0})}{s_0 \cdots s_k} = \sfx{E}{ws_0 \cdots s_k} = \sfx E{(\pi^k)_0}$ and $s_k s_{k+1} \cdots = \mathrm{flat}(\pi^k)$ for all $k \in \N$, the induction hypothesis can be applied to $\varphi_1$ and $\varphi_2$ to conclude the property for $\varphi_1 \,\mathbf U\, \varphi_2$.

	\item ($\Phi$) $\pi \vDash \Phi$ is defined as $\pi_0 \vDash \Phi$ in the standard sense, and $\sfx E{ws_0}, s_0s_1\cdots \vDash \Phi$ is $\sfx E{ws_0} \vDash \Phi$ in the strategy case. Since $\pi_0 = ws_0$, both statements are related as in the induction property. The hypothesis on $\Phi$ itself can be applied, considering that state satisfaction is below path satisfaction in the induction order (we have never used this argument in reverse), and then they are equivalent.

\end{itemize}

	A complete subset of CTL* constructors has been handled in the proof, the derived operators follow from the well-known semantic equivalences.
\end{proof}

\setcounter{proposition}{4}
\begin{proposition}
	Given $(\mathcal K, E)$ and a closed $\mu$-calculus formula $\varphi$, $s \in \mus[\mathcal K, \eta]{\varphi}$ iff $\Gamma_{\mathcal K, s} \in \muss[\mathcal K, \xi]\varphi$ for any $\eta$ and $\xi$.
\end{proposition}

\begin{proof}
This property can be proven inductively, adding the variable valuations to the inductive property and the premise that $\eta(Z) \ni s$ iff $\Gamma_{\mathcal K, s} \in \xi(Z)$ for all variables $Z$. For the initial $\varphi$, this premise is trivially satisfied since we can take $\eta(Z) = \emptyset = \xi(Z)$ regardless of the given two, since the formula is closed. We will not detail some trivial cases:

	\begin{itemize}
		\item ($p$) By definition, $s \in \mus p$ is $p \in \ell(s)$ and  $\Gamma_s \in \muss p$ is $\forall \pi \in \Gamma_s \; p \in \ell(\pi_0)$. Since $\Gamma_s$ are the executions of $\mathcal K$ starting at $s$, $\pi_0 = s$ and both statements are equivalent.
		\item ($\langle a \rangle \varphi$) $s \in \mus{\langle a \rangle \varphi}$ if there is an $s' \in S$ such that $s \to^a s'$ and $s' \in \mus\varphi$. On the other side, $\Gamma_s \in \muss{\langle a \rangle \varphi}$ holds iff there is $saw \in \Gamma_s$ such that $\sfx {\Gamma_s}{saw_0} = \Gamma_{w_0} \in \muss\varphi$. The induction hypothesis with $s' = w_0$ lets us conclude the property.
		\item ($\nu Z . \varphi$) $s \in \mus{\nu Z . \varphi}$ iff there is a set $V$ such that $s \in V$ and $V \subseteq \mus[{\eta[Z/V]}]\varphi$. In the strategy side, $\Gamma_s \in \muss{\nu Z . \varphi}$ iff there is an $F$ such that $\Gamma_s \in F$ and $F \subseteq \muss[{\xi[Z/F]}]{\varphi}$. Assuming there exists a $V$ with these properties ($\Rightarrow$), consider $F = \{ \Gamma_s : s \in V \}$. In other words, $s \in V$ iff $\Gamma_s \in F$, so $\eta[Z/V]$ and $\xi[Z/F]$ are properly related. Hence, by induction hypothesis on $\varphi$, $\Gamma_s \in \muss[{\xi[Z/F]}]{\varphi}$ iff $s \in \mus[{\eta[Z/V]}]{\varphi}$, so $F \subseteq \muss[{\xi[Z/F]}]{\varphi}$ as we wanted to prove. In the opposite direction ($\Leftarrow$), assuming the existence of an $F$ with the mentioned properties, consider $V = \{ s \in S : \Gamma_s \in F \}$ and the proof is the same.
	\end{itemize}
\end{proof}

\begin{proposition}
	Given $(\mathcal K, E(\lambda))$ and a closed $\mu$-calculus formula $\varphi$, $s \in \lBrack \varphi \rBrack_{\mathcal U(\mathcal K, \lambda), \eta}$ for $s\pi \in E$ iff $E \in \lAngle \varphi \rAngle_{\mathcal K, \xi}$ for any $\eta$ and $\xi$.
\end{proposition}

\begin{proof}
	Let us inductively prove the more general property that $\mus\varphi \ni w$ iff $\sfx Ew \in \muss\varphi$ provided that $\eta(Z) \ni w$ iff $\sfx Ew \in \xi(Z)$ for all variables $Z$.
\begin{itemize}
	\item ($p$) By definition, $ws \in \mus\varphi$ iff $p \in \ell(s)$, and $\sfx E{ws} \in \muss\varphi$ iff $p \in \ell(\pi_0)$ for all $\pi \in \sfx E{ws}$. However, $\pi_0$ must be $s$ since $\sfx E{ws} = \{ sw' : wsw' \in E \}$, so both sides are equivalent.
	\item ($Z$) The value of $Z$ in both contexts is respectively $\eta(Z)$ and $\xi(Z)$, so the property directly follows from the assumption about these two functions.
	\item ($\varphi_1 \wedge \varphi_2$) The standard definition says $\mus{\varphi_1 \wedge \varphi_2} = \mus{\varphi_1} \cap \mus{\varphi_2}$ and the strategy one is $\muss{\varphi_1 \wedge \varphi_2} = \muss{\varphi_1} \cap \muss{\varphi_2}$. Hence, the property holds by induction hypothesis on $\varphi_1$ and $\varphi_2$.
	\item ($\neg \varphi$) By definition, $\mus{\neg \varphi} = S^+ \backslash \mus\varphi$ and $\muss{\neg \varphi} = \mathcal P(\Gamma_{\mathcal K}) \backslash \muss\varphi$, so the property holds by induction hypothesis on $\varphi$.
	\item ($\langle a \rangle \varphi$) $ws \in \mus{\langle a \rangle \varphi}$ iff there is an $(a, s') \in \lambda(ws)$ such that $wsas' \in \mus{\varphi}$ according to the standard definition of $\mu$-calculus and the transition relation on $\mathcal U(\mathcal K, \lambda)$. On the other side, $\sfx E{ws} \in \muss{\langle a \rangle \varphi}$ iff there is a $w' \in (S \cup A)^\infty$ such that $saw' \in \sfx E{ws}$ and $\sfx{(\sfx E{ws})}{saw'_0} = \sfx E{wsaw'_0} \in \muss\varphi$.

		By definition of $E(\lambda)$, there is a $w' \in (S \cup A)^\infty$ such that $wsaw' \in E$ iff $(a, w'_0) \in \lambda(ws)$. Hence, by induction hypothesis on $\varphi$ and taking $w'_0 = s'$, we conclude that the property holds.

	\item ($\nu Z . \varphi$) According to the standard definition, $ws \in \mus{\nu Z . \varphi}$ iff there is a $V \subseteq S^+$ such that $V \subseteq \mus[{\eta[Z/V]}]{\varphi}$ and $ws \in V$. According to our definition for strategies, $\sfx E{ws} \in \muss{\nu Z . \varphi}$ iff there is an $F \subseteq \mathcal P(\Gamma_{\mathcal K})$ such that $F \subseteq \muss[{\xi[Z/F]}]{\varphi}$ and $\sfx E{ws} \in F$.
	Both implications can be proven like in the previous proposition, but taking $F = \{ \sfx E{w} : w \in V \}$ for a given $V$, and $V = \{ w \in (S \cup A)^+ : \sfx E{w} \in F \}$ for a given $F$.
\end{itemize}
\end{proof}

\begin{theorem}
	$\ltmsl'_\alpha$ and $\mathcal U(\mathcal M, \lambda_{E(\alpha)})$ are bisimilar Kripke structures.
\end{theorem}

\begin{proof}
Let $f : (T_\Sigma \cup A)^+ \to \mathcal P(\xs)$ be defined by $f(t_0 a_1 t_1 \cdots a_n t_n) = \{ q_n \in \xs : t_0 \ao \alpha = q_0 \opsem^{a_1} \cdots \opsem^{a_n} q_n, \cterm(q_k) = t_k \}$. For any $w \in (T_\Sigma \cup A)^+$ and $Q \neq \emptyset$, $f(w) \, [\opsem] \, Q$ holds if and only if $\exists \, t \in T_\Sigma, a \in A \;\; Q = f(wat)$, since:
\begin{align*}
	f(w) \, [\opsem] \, Q	&\iff \exists \, t \in T_\Sigma, a \in A \;\; Q = \{ q' \in \xs : q \in f(w), q \opsem^a q', \cterm(q') = t \} \\
				&\iff \exists \, t \in T_\Sigma, a \in A \;\; Q = \{ q' \in \xs : t_0 \ao \alpha = q_0 \opsem^{a_1} \cdots \opsem^{a_n} q_n \opsem^a q', \\
				&\kern4.5cm \cterm(q') = t, \cterm(q_k) = w_k \} = f(wat)
\end{align*}

	The relation $R = \{(t_0w, f(t_0w)) : w \in (T_\Sigma \cup A)^*, f(t_0w) \neq \emptyset \}$ is the bisimulation we are looking for. Clearly, $(t_0, \{t_0 \ao \alpha\} ) \in R$ and $\ell_{\mathrm{last}}(ws) = \ell(s) = \ell(\mathrm{cterm}(Q))$ if $(ws, Q) \in R$. Given two words $v, w \in (T_\Sigma \cup A)^+$, $R$ only relates them to $f(v)$ and $f(w)$, respectively. $(v, w) \in U$ implies $w = vat$ by definition of $U$, and then $f(v) \, [\opsem] \, f(w)$ follows from the previous paragraph. Given two non-empty sets $Q$ and $Q'$ such that $Q  \, [\opsem] \, Q'$, and a word $w$ with $f(w) = Q$, we must find a $w'$ such that $(w, w') \in R$ and $f(w') = Q'$. However, we already have it thanks to the previous paragraph and $f(w) \, [\opsem] \, Q'$, since there is some $a$ and $t$ such that $f(wat) = Q'$. It remains to prove that $(w, wat) \in U$, i.e.\ $(a, t) \in \lambda(w)$, but since there are no failed states in $\mathcal M'_\alpha$, any step of the semantics must be allowed by the strategy.
\end{proof}

\bibliographystyle{elsarticle-harv}

\end{document}